\documentclass{itor}

\usepackage{hyperref}
\usepackage{hypcap}
\usepackage{natbib}%
\usepackage[figuresright]{rotating}
\usepackage{mdframed}

\usepackage[T1]{fontenc}
\usepackage[utf8]{inputenc}

\usepackage{amsmath,amssymb,amsfonts,mathtools}

\usepackage{url}
\usepackage[table]{xcolor}
\usepackage{todonotes}
\usepackage{subcaption}

\usepackage{longtable}
\usepackage{pdflscape}
\usepackage{booktabs,multirow}
\usepackage{makecell}
\usepackage{float,placeins}
\usepackage{comment}
\usepackage[toc]{appendix}
\usepackage[normalem]{ulem}

\usepackage{algpseudocode}
\usepackage[linesnumbered,ruled]{algorithm2e}
\SetArgSty{textnormal}
\DontPrintSemicolon

\usepackage{lineno}

\usepackage[tracking=true, protrusion=true, expansion, kerning=true]{microtype}

\usepackage{amsmath,amsthm}

\hypersetup{
	hidelinks
}

\definecolor{lightgray}{gray}{0.9}

\graphicspath{{./}{fig/}}

\begin{document}

\title{\vspace{-72pt}A biased random-key genetic algorithm for the\\home health care problem}

\author[]{
	Alberto F. Kummer\affmark{a},
	Olinto C.B. de Araújo\affmark{b},
	Luciana S. Buriol\affmark{a},
	Mauricio G.C. Resende\affmark{c}
}

\affil{\affmark{a}Federal University of Rio Grande do Sul, Av. Bento Gonçalves, 9500, Porto Alegre, Rio Grande do Sul, Brazil}
\affil{\affmark{b}Federal University of Santa Maria, Av. Roraima, 1000, Santa Maria, Rio Grande do Sul, Brazil}
\affil{\affmark{c}Amazon.com, Inc. and University of Washington, Seattle, Washington, USA.}
\email{afkneto@inf.ufrgs.br; olinto@ctism.ufsm.br;\\buriol@inf.ufrgs.br; mgcr@berkeley.edu}

\normalsize




\begin{abstract}
Home health care problems consist of scheduling visits to home patients by health professionals while following a series of requirements. \color{black} This paper studies the Home Health Care Routing and Scheduling Problem, which comprises a multi-attribute vehicle routing problem with soft time windows. Additional route inter-dependency constraints apply for patients requesting multiple visits, either by simultaneous visits or visits with precedence. We apply a mathematical programming solver to obtain lower bounds for the problem. We also propose a biased random-key genetic algorithm, and we study the effects of additional state-of-art components recently proposed in the literature for this genetic algorithm. \normalcolor We perform computational experiment using a publicly available benchmark dataset. \color{black} Regarding the previous local search-based methods, we find results up to 26.1\% better than those of the literature. We find improvements from around 0.4\% to 6.36\% compared to previous results from a similar genetic algorithm. \normalcolor
\end{abstract}


\maketitle

\section{Introduction}

The increase in life expectancy directly impacts the demand for health services, thus motivating public authorities to design and implement such services. Home Health Care (HHC) was proposed as an affordable alternative to provide health assistance to the population, leveraging hospital bed availability and the comfort and sanitation of home-assisted patients \citep{eveborn2006,frifita2017}. Home health care prevents moving the elderly from their homes, positively impacting their mental health and quality of life \citep{gutierrez2013}. In most cases, home health care is a less intrusive option than nursing homes \citep{eveborn2006}. People currently receiving medical treatments can also benefit from this type of service, which reduces hospitalization stress \citep{landers2016}. It also allows environmental conditions, \color{black}which is important in scenarios such as the COVID-19 pandemic. Thus, a home health provider can implement a protocol to follow-up low-risk patients at home, which helps to reduce the spread of contaminant pathogens. This way, only medium-to-high risk patients would need to receive total healthcare assistance at hospitals.\normalcolor{} 

\newpage

\color{black} According to the literature, the HHC was introduced to increase the access to healthcare services by people living in rural areas \citep{fernandez1974}. \normalcolor{} Nowadays, it is the social component that motivates the implementation of public and private initiatives of HHC. \textcolor{black}{For example, there are initiatives focused on enhancing the population's quality of life by providing services like psychology, occupational therapy, and physiotherapy to encourage the physical exercise and maintain the mental well-being of the elderly \citep{maya2015}}. From a management perspective, home care is the most cost-effective among \textcolor{black}{other} alternatives \citep{gutierrez2013}.

\textcolor{black}{For those reasons, the HHC} has attracted more attention in the last two decades, and we see initiatives to implement services in all developed countries worldwide. The counterpart is that HHC practitioners are faced with the complex logistics of routing caregivers around the coverage area while taking care of fulfilling regulations, maintaining work balance, and maximizing \textcolor{black}{the preferences of both nurses and patients}. The lack of computational tools to optimize HHC operations often leads to poor solutions, incurring problems like unnecessarily costly routing and poor scheduling of caregiver working times \citep{grenouilleau2019}.

\color{black}
This paper discusses efficient solution methods for the daily Home Health Care Routing and Scheduling problem (HHCRSP), introduced by \citet{mankowska2014}.
The contributions of this work can be summarized as follows.

\begin{itemize}
   \item Improved lower bounds by simple tweaks of a standard mathematical optimization software;
   \item A new heuristic-powered decoding strategy for the HHCRSP;
   \item Analysis of state-of-art intensification components of biased random-key genetic algorithms;
   \item Study how more convoluted decoding strategies compare against a simpler decoding algorithm.
\end{itemize}

\normalcolor

The remainder of the paper is organized as follows. Section~\ref{section:previous work} reviews related HHC literature, focusing on the daily HHC problem. Section~\ref{section:the hhcp} defines the problem addressed in this research. Section~\ref{section:proposed-method} introduces our proposed solution method using a biased random key genetic algorithm. This section also discusses the relevant implementation details of our metaheuristic algorithm. In Section~\ref{section:computational results}, we present the outcomes of the proposed solution method, and we perform a comparison with previous results from the literature. Lastly, we summarize the findings of this study and point to some potential future work in Section~\ref{section:conclusion}.

\section{Previous work}
\label{section:previous work}

\color{black}
The HHC problem (HHCP) is not a new research topic in the OR literature, and the first paper dates back to the 1970s with the seminal work of \citet{fernandez1974}. Since then, we have observed an increasing interest in these problems, and their state-of-art is very complex yet robust by considering several practical constraints. Typically, the core feature set of HHCP consists of a routing problem, a set of skilled/qualified caregivers, and a set of patients spread over a geographical region. In the VRP literature, this is often referred to as a multi-attribute Vehicle Routing Problem \citep{grenouilleau2019}. Most authors consider a single-depot problem \citep{cisse2017,grieco2020}, and the Euclidean space is frequently used to model travel times between locations. Recent surveys indicate the prevalence of single transportation modes \citep{fikar2017,grieco2020}. Patients usually have hard time windows \citep{bredstrom2008,rasmussen2012}, but some authors allow some flexibility via soft time windows \citep{mankowska2014,decerle2018}. Any combination of these features configures a complex vehicle routing problem. Furthermore, this is not an exhaustive list of characteristics from the HHC literature, and there is no consensus of what configures a standard problem \citep{fikar2017,grenouilleau2019,grieco2020}.

To better define the scope of our research, we focus our study on the variant of the daily home health care problem introduced by \citet{mankowska2014}. This problem considers multiple service types, caregiver qualification levels, soft time windows, and route inter-dependency constraints for patients requesting multiple visits. \citet{bredstrom2008} approached a similar problem with hard time windows, and they also model caregiver preferences through additional soft constraints. Both publications applied mathematical programming models, solving only small instances to optimality. In both cases, the most promising algorithms were heuristics. \citet{mankowska2014} solved instances with up to 300 patients and 40 caregivers, and \citet{bredstrom2008} solved instances with up to 80 patients and 16 caregivers.

%
%

\citet{lasfargeas2019} proposed stochastic local-search operators to solve HHCRSP of \citet{mankowska2014}. Similar to the literature, the authors used a matrix-based structure to represent solutions and facilitate the implementation of the local search operators, especially the ones for optimizing multiple visits patients. The authors tested their algorithm over the benchmark dataset proposed by \citet{mankowska2014}. According to their results, the stochastic approach benefits the local search technique, leading to improved results and shorter computational times compared to the deterministic algorithms proposed by \citet{mankowska2014}. Despite that, the authors only tested instances up to 75 patients, supposedly due to feasibility issues on larger test cases.

\citet{rasmussen2012} approached a very similar problem to that of \citet{mankowska2014} but considered hard time windows for the patient visits. They modeled the patient assignment as soft constraints to circumvent feasibility issues, allowing a solution to have unscheduled visits. The authors also generalized the route inter-dependency requirements for multiple visit patients, and they propose a master formulation using such generalized constraints \citep{desaulniers2006}. They applied a branch-and-price algorithm, but the generalized constraints were implicitly enforced during the branching phase, so there is no need to develop a specialized algorithm for the pricing problem. Proposed algorithms were tested over the dataset of \citet{bredstrom2008}. The authors also devised new random instances, partially based on a real HHC problem arriving in Denmark. They solved instances of 15 caregivers and 150 patients exactly with a branch-and-price algorithm, exploiting the problem structure with tailored branching rules from routing problems with transshipment \citep{drexl2012}.

\normalcolor

A few other authors further tested the dataset of \citet{mankowska2014}. \citet{neto2019} provide new solutions via a fix-and-optimize \textit{matheuristic} that, iteratively, efficiently re-optimizes small portions of the problem. \textcolor{black}{They applied the constructive heuristic of \citet{mankowska2014} to generate initial solutions,} and the \textit{matheuristic} \textcolor{black}{runs until achieving a certain number of non-improving iterations.} The authors were capable of improving the best-known solutions to instances of up to 100 patients in a competitive computational time compared to \citet{mankowska2014} and \citet{lasfargeas2019}. Their fix-and-optimize demonstrated to be inapplicable in instances with more than 100 patients, \textcolor{black}{mainly due to high memory consumption by the solver.}

\color{black}
\citet{kummer2020} were the first to propose a biased random-key genetic algorithm (BRKGA) \citep{gonccalves2011art} for solving an HHC problem. Their decoder employed a greedy best-insertion heuristic that selects the best caregiver(s) to fulfill patient demands, and the patients were processed according to the permutation derived from the chromosome. Their metaheuristic algorithm was tested on the instances from \citet{mankowska2014}. According to the authors, the exploit-and-explore capabilities of their BRKGA performed consistently better than the local search operators from the literature, mainly in medium to large-sized instances. Due to the greediness of the decoder used, the metaheuristic \textcolor{black}{algorithm} could not achieve \textcolor{black}{an} optimal solution for some small instances. Despite that, the proposed algorithm outperformed almost all previous literature results for instances with 25 patients and five caregivers up to the largest test cases of the dataset, comprising 300 patients and 40 caregivers.

Considering that there is still no agreement in the literature about a standard home health care problem, we summarize in Table~\ref{table:previous-works} the closely related literature to the single-day HHC problem we study. For a more comprehensive overview of the literature, please c.f. \citet{cisse2017}, \citet{fikar2017}, and \citet{grieco2020}.
\normalcolor

\begin{table}[!htb]
	\centering
	\caption{Summary of publications regarding the HHCP.}
	\label{table:previous-works}
	\bgroup
	\scriptsize
	\setlength{\tabcolsep}{3pt}
	\renewcommand{\arraystretch}{1.2}
   \color{black}
	\begin{tabular}{lccccccccccccccccccc}
		\toprule
		\multirow{2}[2]{*}{Work} &
		\multirow{2}[2]{*}{M} &
		\multirow{2}[2]{*}{PH} &
		\multirow{2}[2]{*}{RM} &
		\multirow{2}[2]{*}{T} &
		\multirow{2}[2]{*}{WT} &
		\multirow{2}[2]{*}{PR} &
		\multirow{2}[2]{*}{PTW} &
		\multirow{2}[2]{*}{CTW} &
		\multirow{2}[2]{*}{WB} &
		\multirow{2}[2]{*}{WT} &
		\multirow{2}[2]{*}{OT} &
		\multirow{2}[2]{*}{BR} &
		\multirow{2}[2]{*}{SK} &
		\multirow{2}[2]{*}{CP} &
		\multirow{2}[2]{*}{SY} &
		\multirow{2}[2]{*}{SM} &
		\multicolumn{3}{c}{Largest inst. solved}\\
		\cmidrule{18-20}
		&&&&&&&&&&&&&&&&& \#C & \#P & \#J\\
		\midrule
		\citet{fernandez1974} & -- & S & TD & H & -- & -- & -- & -- & -- & -- & -- & -- & -- & -- & -- & -- & -- & -- &    \\
		\citet{bredstrom2008}$^\dag$ & x & S & TT & H & -- & S & H & H & S & -- & -- & -- & -- & -- & S, P & MIP+RH & 16 & 80 & 80 \\
		\citet{rasmussen2012}$^\dag$ & x & S & TC & S & -- & S & H & H & -- & -- & -- & -- & H & -- & S, P & B\&P & 16 & 150 & 150 \\[8pt]
		\citet{mankowska2014}$^\ast$ & x & S & TT & H & -- & -- & S & -- & -- & -- & -- & -- & H & -- & S, P & \makecell{MIP, \\AVNS} & 40 & 300 & 390 \\[8pt]
		\citet{lasfargeas2019}$^\ast$ & x & M & TT & H & -- & -- & S & H & -- & -- & -- & -- & H & H & S, P & VNS & 15 & 75 & 98 \\
		\citet{decerle2018}$^\dag$ & x & S & TD & H & -- & -- & S  & H & -- & -- & -- & -- & H & -- & S & MA & 6 & 109 & 109 \\
		\citet{neto2019}$^\ast$ & x & S & TT & H & -- & -- & S & -- & -- & -- & -- & -- & H & -- & S, P & F\&O & 20 & 100 & 130 \\
		\citet{kummer2020}$^\ast$ & x & S & TT & H & -- & -- & S & -- & -- & -- & -- & -- & H & -- & S, P & GA & 40 & 300 & 390 \\
		\bottomrule
	\end{tabular}
	\caption*{\scriptsize \textbf{M:} Proposes a model to the problem; \textbf{PH:} Length of the planning horizon (S: short/single day, M: medium/up to one week, L: long/several months); \textbf{RM:} Routing metric used to evaluate the solution cost of the routing (TD: travel distance, TC: travel cost, TT: travel time); \textbf{T:} Tasks must be fulfilled (H: hard) or may not (S: soft, with penalization in the objective function); \textbf{WT:} Constraints to model waiting times; \textbf{PR:} Preference of assignments between caregivers and patients; \textbf{PTW:} Patient time-windows; \textbf{CTW:} Caregiver time-windows; \textbf{WB:} Working balance constraints; \textbf{WT:} Constraints to model work time regulations; \textbf{OT:} Constraints to model overtime; \textbf{BR:} Breaks/rest time constraints; \textbf{SK:} Presence of skills/qualification levels for the caregivers; \textbf{SY:} Route inter-dependency constraints (S: simultaneous attendance, P: precedence constraints); \textbf{SM:} \textcolor{black}{Best-performing} solution method (LP+RH: linear programming plus rounding heuristic, RMH: repeated matching heuristic; PSO: particle-swarm optimization, MIP+RH: MIP with rounding heuristic, VNS: variable neighborhood search, B\&P: branch-and-price solver, AVNS: adaptive VNS, MA: memetic algorithm, F\&O: fix-and-optimize matheuristic, GA: genetic algorithm); \textbf{\#C:} Number of caregivers; \textbf{\#P:} Number of patients; \textbf{\#J:} Number of jobs. \textcolor{black}{Works that use the benchmark dataset proposed in \citet{bredstrom2008}$^\dag$, and \citet{mankowska2014}$^\ast$, respectively .}}
	\egroup
\end{table}

\section{The home health care problem}
\label{section:the hhcp}

\citet{mankowska2014} proposed a single-day problem called the home health care routing and scheduling problem (HHCRSP). We use the terms HHCRSP and HHCP interchangeably in this section and the following. Let $\mathcal{V}$ be the set of the caregivers/vehicles, and $\mathcal{C}$ the set of patients/nodes. This routing component aims to assign caregivers to the patients, seeking to minimize each caregiver's travel time. Let $\mathcal{C}^0 = \{0\} \cup \mathcal{C}$ , where $0$ represents the depot. Parameter $d_{ij}$ defines the travel time between the tuples $(i,j) \in \mathcal{C}^0 \times \mathcal{C}^0$. All caregivers travel at the same time--i.e. the problem considers a single transportation modality. The patients also have time-windows. More precisely, each patient $i \in \mathcal{C}$ has a hard time-window starting $e_i$, meaning that a visit can only happen after the start of the time-window. Conversely, patient $i$ also has a soft time-window ending $l_i$, and such services should be served ideally before the end of the patient time-window. If such a requirement is not fulfilled, then attendance is considered tardy, and a penalty is incurred in the objective function. In the next section, we introduced service types, a characteristic of this HHCP that sets routing components as multi-attribute VRPTW \citep{grenouilleau2019}.

\subsection{Service types, qualification, and job requirements}
\label{subsection:skills}

In the HHCRSP, a patient may request up to two distinct \textit{service types}, which skilled caregivers must fulfill. This feature is inherent to the HHCP, requiring multiple visits to some patients. Let $\mathcal{S}$ be the set of service types considered in the problem. Caregivers are skilled in the sense that each caregiver has a subset of service types that they can perform. For each $v \in \mathcal{V}$ and service type $s \in \mathcal{S}$, the parameter $a_{vs} = 1$ indicates if caregiver $v$ can perform service type $s$, and is otherwise 0. Patients have a similar parameter setting. The parameter $r_{is} = 1$ indicates if patient $i \in \mathcal{C}$ requires attendance for service type $s \in \mathcal{S}$, and is otherwise 0. Additionally, the processing time required to complete such attendance is defined according to the parameter $p_{is} > 0$.

In general, the problems are composed mostly of \textit{single service} patients, requiring a visit by a single skilled caregiver. Patients requesting two distinct service types are known as \textit{double service} patients. We introduce the set $\mathcal{C}^\text{d} \subseteq \mathcal{C}$ to refer to the subset of double service patients. In double services, two visits are required by distinct (and skilled) caregivers to fulfill each patient's demands, and additional constraints are imposed when performing such visits.

\subsection{Synchronization constraints for double service patients}
\label{subsection:synchronization}

\begin{figure}[!htb]
	\centering
	\subfloat[Double service with simultaneous attendance]{
		\includegraphics[width=0.48\linewidth,page=2]{sync-tsn2.pdf}
		\label{figure:sync-simult}
	}
	\hfill
	\subfloat[Double service with precendece constraint]{
		\includegraphics[width=0.48\linewidth,page=1]{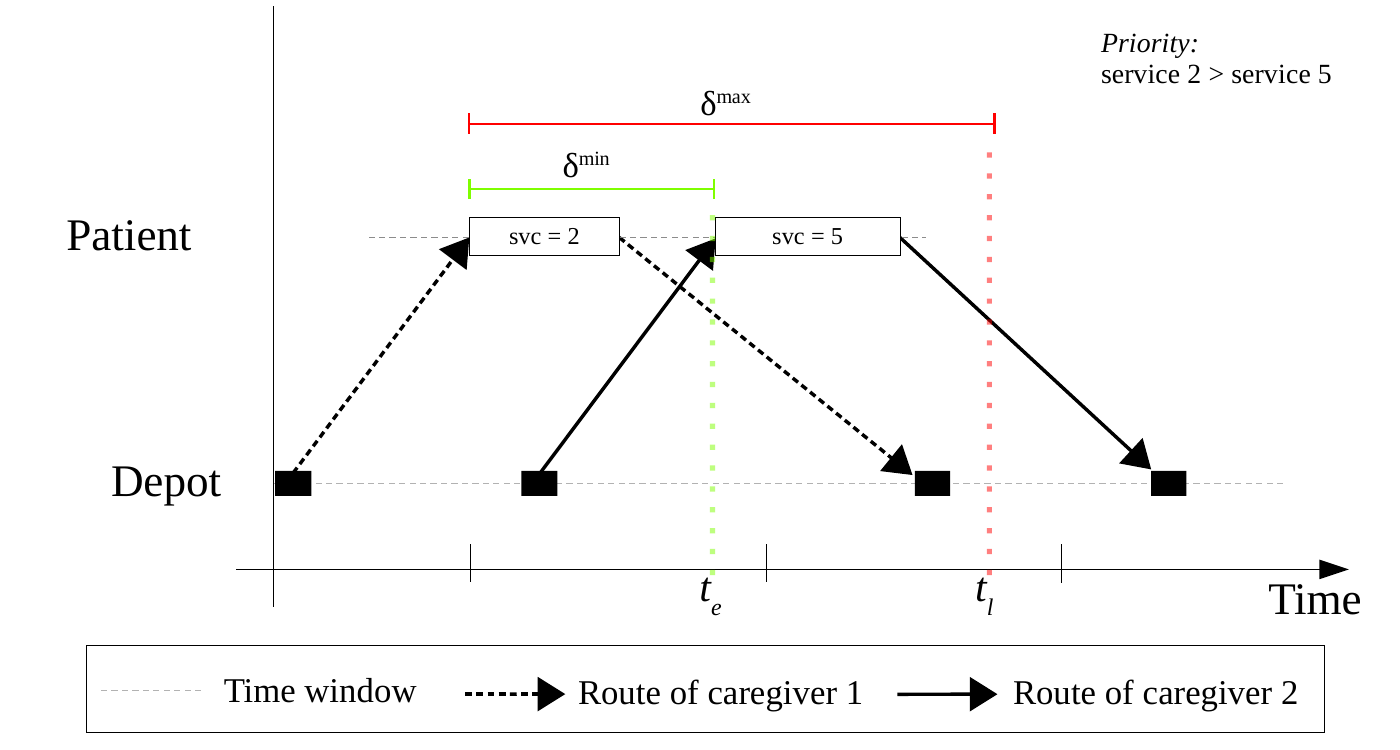}
		\label{figure:sync-pred}
	}
	\caption{A time-space visualization of a double service patient. In \ref{figure:sync-simult}, caregiver \#1 arrives at patient \#1 before caregiver \#2, thus the attendance of the services are delayed until both caregivers are present with patient \#1. In \ref{figure:sync-pred}, caregiver \#2 must start service type \#5 within the \textit{dynamic time-window} $\left[t_e, t_l\right]$.}
\end{figure}

As aforementioned, some patients require more than one service type. For those, two new constraints, inherent to the scope of HHCP, are imposed. These constraints establish route inter-dependency among the caregivers, either in \textit{simultaneous attendances} or by \textit{precedence constraints}. A first requirement is that both caregivers be simultaneously present with the patient. Thus, the processing of both services starts simultaneously. Additionally, the caregivers can only depart from the patient after the completion of the longest of the two services, as shown in Figure \ref{figure:sync-simult}.

In precedence constraints, we have a priority order for which the service types requested must be serviced. Without loss of generality, HHCRSP considers a \textit{global priority} among the service types in $\mathcal{S}$. For example, with a set $\mathcal{S} = \{1,2,3\}$, service 1 has priority over services 2 and 3. Similarly, service 2 also has priority over 3. In other words, service $i$ has priority over service $j$ if $i,j \in \mathcal{S}$.

In addition to this priority, patients with precedence constraints require a minimum and maximum separation time between their two service requests. Parameter $\delta^\text{min}$ indicates the minimum difference between the start of both service types, which may cause the caregiver responsible for the lower priority service to wait until the start of their operation. Similarly, parameter $\delta^\text{max}$ indicates the maximum difference between the start time of the services. In this case, such a requirement can cause the caregiver responsible for the high priority service to delay their operation to maintain the solution feasibility. In the example of Figure~\ref{figure:sync-pred}, service type 5 can be started at any time after $t_e$, and before $t_l$.

\subsection{Mixed integer programming model}
\label{subsection:mip}

In addition to discussing the features of the HHCRSP,
we also introduce the model of \citet{mankowska2014} to define the problem formally. This MIP considers three sets of variables. The decision variables $x_{ijvs} \in \{0,1\}$ indicates if caregiver $v \in \mathcal{V}$ departs from $i \in \mathcal{C}^0$ to $j \in \mathcal{C}^0$ to perform the service type $s \in S$ in the destination node. Considering caregiver capabilities and the patient requirements, we can rewrite the bounds of these variables as $x_{ijvs} \in \{0,a_{vs}r_{is}\}$, allowing the removal of several decision variables from the problem. Variable $t_{ivs} \geqslant 0$ indicates the service start time of patient $i \in \mathcal{C}$ by caregiver $v \in \mathcal{V}$ for service type $s \in \mathcal{S}$. Variable $z_{is} \geqslant 0$ is the tardiness when the service type $s \in \mathcal{S}$ starts after the end of the patient's time window.

The HHCP has three minimization criteria, computed in variables $D, T, T^\mathrm{max} \geqslant 0$, respectively, regarding the total travel distances, the total tardiness, and the largest tardiness over all the patients. Those criteria are weighted according the multiplicative factors $\lambda_1$, $\lambda_2$, and $\lambda_3$ in a single objective function. The MIP model for the HHCRSP is given in (\ref{model_obj}--\ref{model_dtt_dom}).
\begingroup
\allowdisplaybreaks
\begin{align}
   \text{Minimize} \quad  \lambda_1 D + \lambda_2 T + \lambda_3 T^\mathrm{max} \label{model_obj}
\end{align}
Subject to:
\begin{align}
   %
   %
   & D = \sum_{v \in \mathcal{V}} \sum_{i \in \mathcal{C}^0} \sum_{j \in
      \mathcal{C}^0} \sum_{s \in \mathcal{S}} d_{ij} x_{ijvs} \label{model_dvar}\\
   %
   %
   & T = \sum_{i \in \mathcal{C}} \sum_{s \in \mathcal{S}} z_{is} \label{model_tvar}\\[8pt]
   %
   %
   & T^\mathrm{\,max} \geqslant z_{is} & & \forall i \in \mathcal{C}, s \in \mathcal{S} \label{model_tmaxvar}\\[8pt]
   %
   %
   & \sum_{i \in \mathcal{C}^0} \sum_{s \in \mathcal{S}} x_{0ivs} =
   \sum_{i \in \mathcal{C}^0} \sum_{s \in \mathcal{S}} x_{i0vs} & & \forall v \in \mathcal{V} \label{model_src_sink}\\
   %
   %
   & \sum_{j \in \mathcal{C}^0} \sum_{s \in \mathcal{S}} x_{jivs} = \sum_{j \in \mathcal{C}^0}
   \sum_{s \in \mathcal{S}} x_{ijvs} & & \forall i \in \mathcal{C}, v \in \mathcal{V} \label{model_flow_conservation}\\
   %
   %
   & \sum_{v \in \mathcal{V}} \sum_{j \in \mathcal{C}^0} a_{vs}
   x_{jivs} = r_{is} & & \forall i \in \mathcal{C}, s \in \mathcal{S} \label{model_assignment}\\[4pt]
   %
   %
   & t_{ivs_1} + p_{is_1} + \mathrm{d}_{ij} \leqslant t_{jvs_2} + M \left( 1-x_{ijvs_2} \right)
   &
   \begin{split}
   & \forall i \in \mathcal{C}^0, j \in \mathcal{C},\\
   & v \in \mathcal{V}, s_1, s_2 \in \mathcal{S}
   \end{split} \label{model_subcycle_elim}\\[4pt]
   %
   %
   & t_{ivs} \geqslant e_i &
   & \forall i \in \mathcal{C}, v \in \mathcal{V}, s \in \mathcal{S} \label{model_tw_start}\\
   %
   %
   & t_{ivs} \leqslant l_i + z_{is} &
   \begin{split}
   & \forall i \in \mathcal{C}, v \in \mathcal{V}, \\
   & s \in \mathcal{S}
   \end{split}\label{model_tw_end}\\[11pt]
   %
   %
   \begin{split}
   & t_{i v_2 s_2} - t_{i v_1 s_1} \geqslant \delta^\mathrm{min}_i \\
   & \qquad - M \left( 2 - \sum_{j \in \mathcal{C}^0} x_{ji v_1 s_1} - \sum_{j \in \mathcal{C}^0} x_{ji v_2 s_2} \right)
   \end{split}
   &
   \begin{split}
   & \forall i \in \mathcal{C}^\mathrm{d}\\
   & v_1, v_2 \in \mathcal{V}, \\
   & s_1, s_2 \in \mathcal{S}: s_1 < s_2
   \end{split} \label{model_sync1}\\[11pt]
   %
   %
   \begin{split}
   & t_{i v_2 s_2} - t_{i v_1 s_1} \leqslant \delta^\mathrm{max}_i \\
   & \qquad + M \left( 2 - \sum_{j \in \mathcal{C}^0} x_{ji v_1 s_1} - \sum_{j \in \mathcal{C}^0} x_{ji v_2 s_2} \right)
   \end{split}
   &
   \begin{split}
   & \forall i \in \mathcal{C}^\mathrm{d},\\
   & v_1, v_2 \in \mathcal{V}, \\
   & s_1, s_2 \in \mathcal{S}: s_1 < s_2
   \end{split} \label{model_sync2}\\[11pt]
   %
   %
   & x_{ijvs} \in \{0, a_{vs} r_{js}\} &
   & \forall i,j \in \mathcal{C}^0, v \in \mathcal{V}, s \in \mathcal{S} \label{model_x_dom}\\
   & t_{ivs},\, z_{is} \geqslant 0 &
   & \forall i \in \mathcal{C}^0, v \in \mathcal{V}, s \in \mathcal{S} \label{model_t_z_dom}\\
   & D,\, T,\, T^\mathrm{max} \geqslant 0 \label{model_dtt_dom}
\end{align}
\endgroup

Expression (\ref{model_obj}) defines the objective function, using the solution quality indicators calculated in constraints~(\ref{model_dvar}--\ref{model_tmaxvar}). Constraints~(\ref{model_src_sink}) indicate that each vehicle should depart and return to the depot once, and constraints (\ref{model_flow_conservation}) model the flow balance to the other vertices. Constraints (\ref{model_assignment}) guarantee that the service requests of the patients are performed by skilled caregivers. Constraints (\ref{model_subcycle_elim}) are the sub-tour elimination constraints considering the distances between nodes and service processing times. Parameter $M$ is a large number. Constraints (\ref{model_tw_start}, \ref{model_tw_end}) model the time windows and compute the services start times on patients. Constraints (\ref{model_sync1}, \ref{model_sync2}) model both simultaneous double services and separation times on double services with precedence. Finally, constraints (\ref{model_x_dom}--\ref{model_dtt_dom}) define the domain of the decision variables.

\section{Proposed solution method}
\label{section:proposed-method}

The weak linear relaxation of model (\ref{model_obj}--\ref{model_dtt_dom}) severely hurts the performance of branch-and-cut methods from the state-of-art MIP solvers \citep{mankowska2014}. Although new valid inequalities can potentially help to strengthen the linear relaxation of the MIP, the number of variables required to model the problem is still an issue, rendering the employment of solvers impractical for instances with more than 100 patients with a standard computer of 8 GB of memory \citep{neto2019}. Furthermore, methods based on the modification of solutions are often ineffective in solving the HHCRSP. Regarding the optimization landscape, the presence of time-windows makes it hard to escape from local optima \citep{gendreau2010}. Previous work in the literature show how complex it is to keep the solution structure for trajectory-based methods in the presence of route inter-dependency constraints \citep{haddadene2016}. \citet{drexl2012} suggests that it may be worthwhile to try to solve problems with route inter-dependencies using an indirect approach over an \textit{auxiliary search space} to overcome feasibility issues.

Instead of crafting nifty local search operators to navigate through the neighborhoods of a promising solution, we propose exploring the problem solution space indirectly through a multi-population multi-parent biased random-key genetic algorithm, with an additional component of implicit path-relinking for intensification of the search \citep{andrade2021}.

\subsection{The biased random-key genetic algorithm}
\label{subsection:brkga}

The biased random-key genetic algorithm (BRKGA) has been applied to a wide variety of combinatorial optimization problems, including the container loading problem, the parallel machines scheduling problem, and project scheduling problems \citep{gonccalves2011,gonccalves2012,chaves2016}. To the best of our knowledge, just one paper in the literature proposes solving the HHCP with a genetic algorithm \citep{kummer2020}.

In a BRKGA, the initial population is comprised of vectors of real-valued random keys from $[0,1)$. Let $p$ be the population size of the genetic algorithm. In each generation, individuals of the current population are sorted according to their fitness values. The best $p_e$ individuals of the population form the elite set, and the other $p - p_e$ individuals compose the non-elite set. The BRKGA is elitist and copies all individuals of the elite set of the current population to the next population. This way, the BRKGA monotonically keeps the best solutions found over the search \citep{gonccalves2011art}.

\begin{figure}[!htb]
	\centering
	\includegraphics[width=0.55\textwidth]{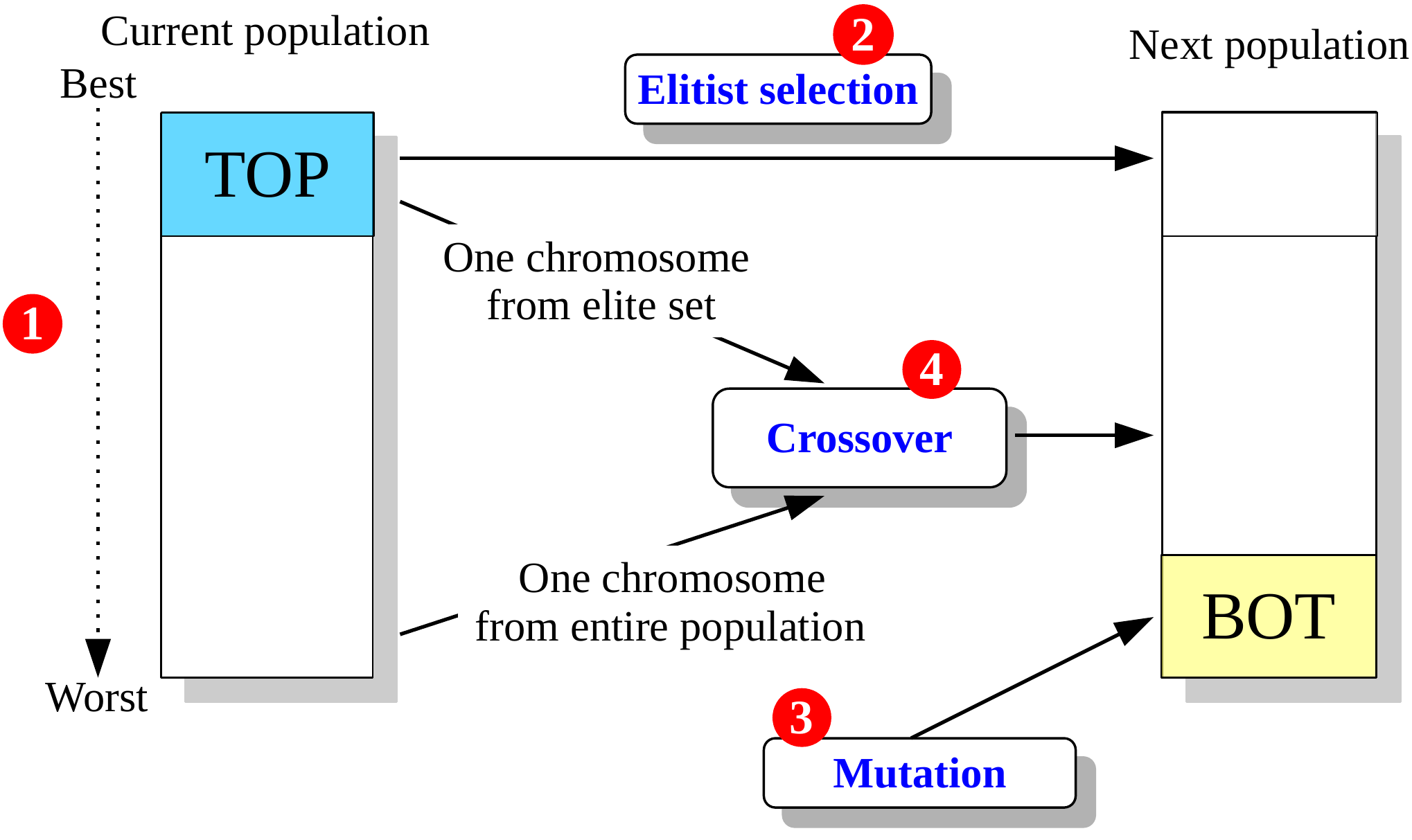}
	\caption{Overview of the BRKGA internals. The operation (1) sorts the current population. In (2) the elite individuals are copied to the next population. In (3), some new mutant individuals are generated to keep the diversity of the next population. In (4), the remaining space of the next population is filled up with offspring.}
	\label{figure:brkga-framewk}
\end{figure}

A BRKGA uses two strategies to create the offspring in the next population. The first strategy aims to keep the new population's diversity by inserting $p_m$ new random mutants in the next population. The second strategy consists of mating one member from the elite set with another member from the non-elite set by employing a parametrized uniform crossover operator to generate offspring. Each allele of the offspring is either inherited from the elite parent, with a probability of $\rho_e$ , or from the non-elite parent, with a probability of $1-\rho_e$. The mating generates $p - p_e - p_m$ new individuals, which are inserted in the next population. Finally, the next population becomes the current population, and the process repeats until achieving a maximum number of generations or some other stopping criterion. Figure~\ref{figure:brkga-framewk} depicts, \textcolor{black}{at a} high level, these operations, highlighting the sequence of operations.

The multi-population BRKGA consists of evolving multiple islands simultaneously. The islands consist of several populations evolving independently, allowing distinct islands to (possibly) achieve distinct local minima. \textit{Immigration} allows exploiting the problem's solution space by moving some individuals from one island to the other. With $k$ individual populations, $m$ immigrants from the island \textcolor{black}{$i$} are moved to the island $\min\{i+1,1\}$, and this process repeats for $i = 1, 2, \dots, k$. Typically, the immigration mechanism is triggered by some criteria during the evolutionary process, e.g., periodically after a fixed number of generations \citep{toso2015}.

A BRKGA has some strengths that help to manage the complicating constraints of the HHCP. It uses a parameterized uniform crossover, dismissing the necessity of repair operators by delegating feasibility issues to the problem's objective function, requiring a well-designed decoder to prevent the evolutionary process from getting stuck due to infeasibilities \citep{gonccalves2011art}. The choice of chromosome representations with random keys and the constant generation of mutants enables exploring the solution space. In contrast, the crossover operator and the decoder exploit the structural properties of the solutions \citep{eiben2003}. Such characteristics also relieve the necessity to employ local search, which is frequently a very time-consuming component of the heuristics and has a negligible impact on the solutions \citep{drexl2012}.

\subsection{Multi-parent BRKGA}
\label{subsection:multi-parent}

As aforementioned, the \textit{standard} BRKGA performs the mating process over two parents from the population.
Further researchers identified simple modifications that improve the convergence of the mating process. Computational experiments presented in \cite{lucena2014-gd-mp} showed that the multi-parent mating produces consistently better results than the standard and the \textit{gender-defining} variant of BRKGA \citep{lucena2014-gd-mp}.

In multi-parent BRKGA, the metaheuristic selects a total of $\pi_t$ parents for each mating, with $\pi_e$ individuals from the elite set, and the other $\pi_t - \pi_e$ from the non-elite set. Contrary to specifying the bias for selecting alleles from the parents directly, the multi-parent BRKGA uses a two-step approach. First, the parents are put into a list, which is then sorted according to their \textit{fitness}. The second step consists of applying a bias function to compute each parent's weight according to their \textit{rank} $r$ in the sorted list. \citet{andrade2021} suggest using one of the non-decreasing bias functions $\Phi$ to compute such weights \citep{bresina1996-bias}.
In addition to these functions, we \textcolor{black}{derived} the quadratic $\Phi(r) = r^{-2}$ and cubic $\Phi(r) = r^{-3}$ functions.

\subsection{Implicit path-relinking in random-keys space}
\label{subsection:implicit-pr}

Path-relinking is an intensification \textcolor{black}{procedure} that explores the intermediate solutions between a base and a guide solution by incrementally introducing changes \textcolor{black}{in the base solution according to the guide}. \textcolor{black}{It finishes when both guide and base solutions are equal \citep{glover1997-pr}, or after exploring a maximum number of intermediate solutions \citep{resende2010scatter}. According to the studies of \citet{resende2016-pr}, the \textit{direction} of this exploration matters.} In the \textit{forward path-relinking}, the changes come from the guide solution. Conversely, the algorithm is called \textit{backward path-relinking} when the changes come from the base solution. \citet{resende2016-pr} also propose the \textit{mixed path-relinking}, which combines both forward and backward strategies. According to their experiments, this mixed intensification approach often finds a better solution than the others.

A typical implementation of the path-relinking heuristic is usually tied closely to the problem \citep{andrade2021}. The heuristic is structured around the solution components (or variables), and how these components can be interchanged between the base and guide solution to generate new intermediate solutions. It is very hard to generalize a single implementation of path-relinking to a wide variety of problems. Nevertheless, \citet{andrade2021} propose a path-relinking implementation that takes advantage of the generic solution encoding of the BRKGA to implement what they call an implicit path-relinking procedure (IPR).

\color{black}
As the reader may expect, there is a significant overhead of applying a path-relinking heuristic inside another metaheuristic such as the BRKGA. For this reason, \citet{andrade2021} also propose a minimum distance metric between the base and the guide solution to allow running the IPR. This strategy relies on the fact that measuring such distances is much faster than running the heuristic itself, so it works as an effective speedup strategy to skip IPR runs between solutions that are too similar.
\normalcolor

\begin{figure}[ht]
	\centering
	\includegraphics[width=0.45\linewidth,page=2]{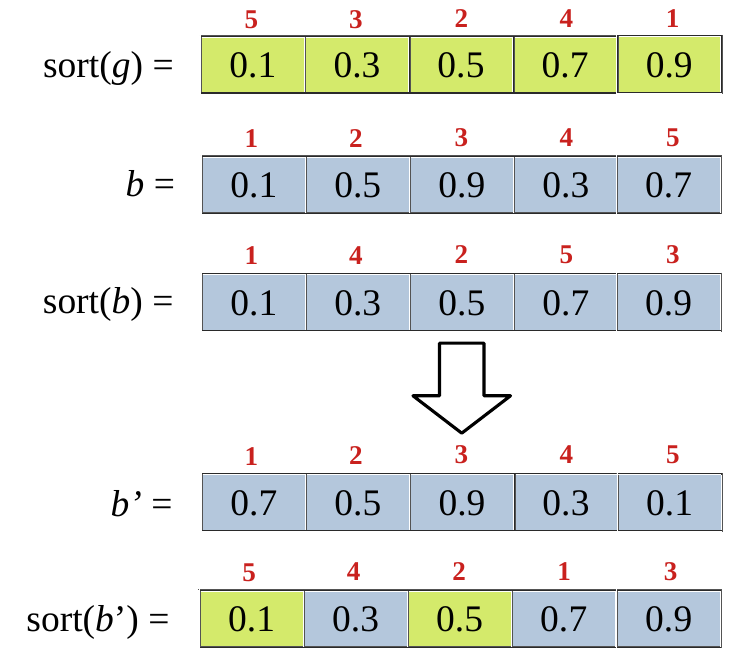}
	\caption{Example of \textit{permutation IPR}. The guide individual $g$ is used to induce changes in the base individual $b$.}
   \label{figure:permutation-ipr}
\end{figure}

\textcolor{black}{\citet{andrade2021} suggested two variations of the IPR} that operates differently according to how the solutions are encoded.
\textcolor{black}{Specifically to} routing and sequencing problems, the decoder uses the random keys to establish a permutation of the solution elements. For a vehicle routing problem, e.g., each client is associated with a chromosome position; thus, the visit order is obtained by sorting the chromosome \citep{ruiz2019}. For this type of solution encoding,
\textcolor{black}{\citet{andrade2021} proposed a \textit{permutation IPR}} that uses the guide chromosome to \textcolor{black}{\textit{induce permutations}} in the base chromosome, as shown in Figure~\ref{figure:permutation-ipr}.
\textcolor{black}{This variation of the implicit IPR starts by sorting both \textit{guide} and \textit{base individuals}. At this point the algorithm tests for the minimal distance criterion, using the Kendall-tau distance over the permutation of array indices of $\mathit{sort}(b)$ and $\mathit{sort}(g)$.
After that, the IPR generates \textit{intermediate individuals} by swapping keys in either \textit{base} or \textit{guide} individual, following a hybrid forward-backward path relinking strategy.
In the example of the figure, the guide individual sets that the first key must be associated with index 5 (according to sort($g$)), but the base solution associates the index 1 to the first key (according to sort($b$)). The IPR then swaps the key associated with the index 5 in $b$ (following the permutation of $g$), with the key currently occupying position $1$ of $b$. This procedure generates the individual $b'$, whose first element (in \textit{sort}($b'$)) respects the first element in the permutation of $g$. After introducing this change, the decoder evaluates the intermediate individual, hopefully finding an improved solution.}

\textcolor{black}{In presence of multiple islands,} the implicit path-relinking is applied between pairs of elite individuals from distinct islands, either by selecting the best individuals or selecting random individuals from \textcolor{black}{their elite sets, following the same circular strategy of the immigration mechanism}. For example, with three islands, the IPR runs between some elite individuals from islands 1 and 2, then from islands 2 and 3, and finally 3 and 1. \textcolor{black}{With the ``best selection'' strategy, each IPR run selects the elite individuals according to their fitness.} With the random selection, the IPR \textcolor{black}{simply} selects two random individuals from the elite sets. \textcolor{black}{When only a single island is considered, the heuristic operates between elite individuals from the same island  \citep{andrade2021}.}

\subsection{Heuristic-powered decoder for the HHCRSP}
\label{subsection:decoding}

\begin{figure}[!htbp]
   \centering
   \includegraphics[width=0.7\textwidth]{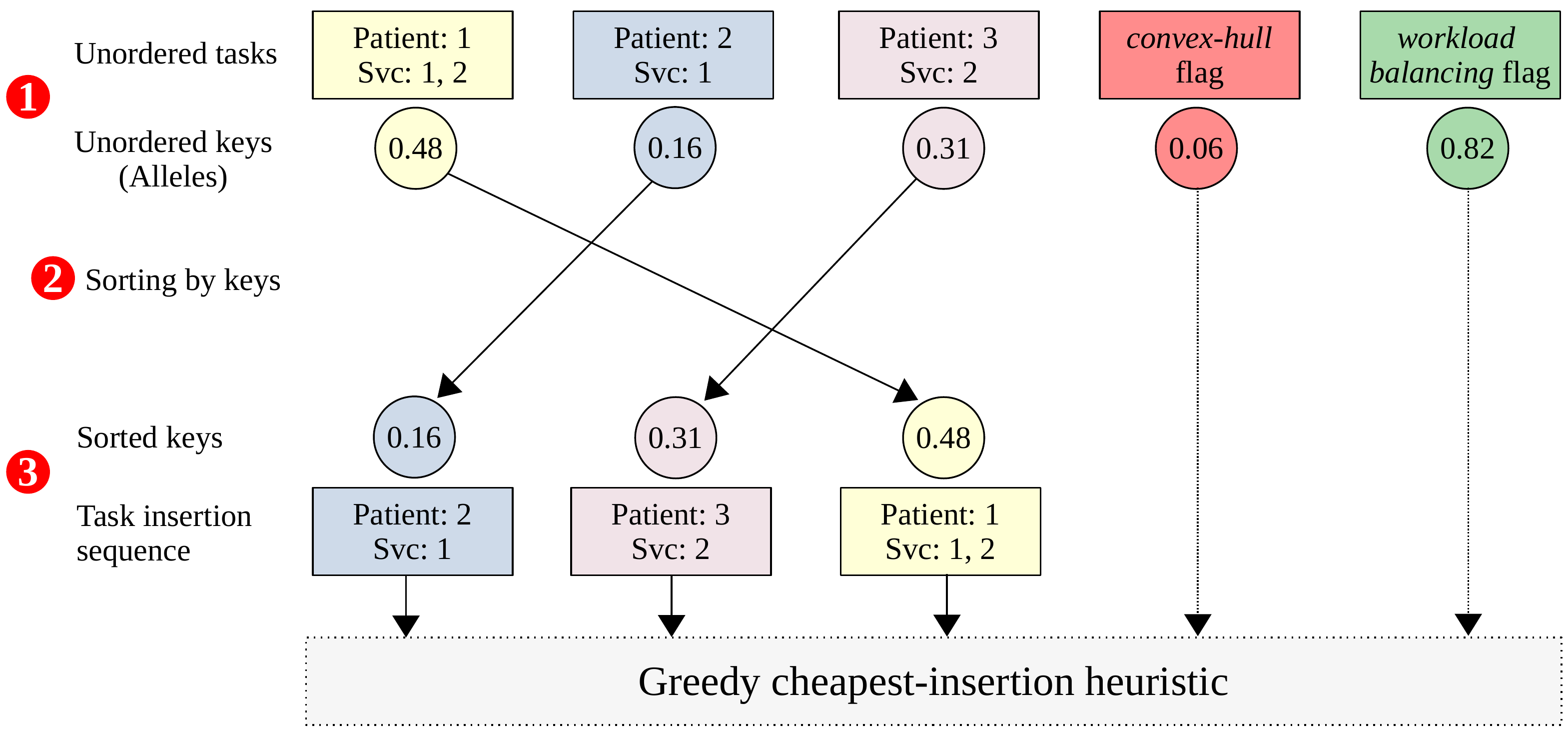}
   \caption{\color{black} Decoding example for a instance with three patients.The greedy phase has the \textit{convex hull} strategy set off and the caregiver workload balancing heuristic set on.}
   \label{figure:decoding-example}
\end{figure}

\color{black}
In our solution approach, the genetic algorithm evolves the sequence in which the patients are inserted into the solution, and each chromosome has length $|\mathcal{C}|+2$. \normalcolor{} The example of Figure \ref{figure:decoding-example} demonstrates the decoding process on an instance with three patients. The algorithm starts pairing the list of patients with the chromosome, then it sorts the task list according to the increasing order of the allele values, reaching the task sequence depicted in~\textit{task insertion sequence}. \textcolor{black}{This process uses the first $|\mathcal C|$ alleles of the chromosome.} Then, a greedy cheapest-insertion heuristic assigns the sorted sequence of tasks to the caregivers, \textcolor{black}{and the last two alleles are used to toggle optional components of the decoding algorithm. When allele $|\mathcal C|+1 < 0.5$, the decoder only considers the increase in total travel time by inserting the patient into the caregiver's route. When this allele has a value $\geqslant 0.5$, then the decoder also includes the distance for returning the caregiver to the depot after servicing the patient. In the context of this manuscript, we refer to this behavior as the \textit{convex hull} strategy. The last allele $|\mathcal C|+2$ toggles a simple heuristic that tries to balance the workload of caregivers during the greedy construction phase. }

\color{black}
Algorithm~\ref{algorithm:greedy-constructive} details the implementation of the proposed decoder. In line 1, the decoder starts an empty solution, with all the vehicles placed at the depot. Line 2 performs the same steps depicted in the example of Figure~\ref{figure:decoding-example}. Lines 3 and 4 check if the optional components should be enabled during the rest of the decoding. Line 5 initializes the total workload of each caregiver as 0. The decoder then takes each task from the sorted task sequence (line 6) and tests all the possible caregiver assignments to such task (lines 7--19), considering all the service types requested by the patient. This is done by passing the service types requested by $\mathit{tasks}\left[i\right]$ to the function \texttt{AllVehicleCombinations}, which returns a list with all pairs of vehicles with qualification to perform the service types $s_1$ and $s_2$, respectively. The heuristic then evaluates the cost of assigning the task to the current vehicles $v_1$ and $v_2$ through \texttt{FindInsertionCost} (line 10). Lines 11--13 update the cost $\bar{c}$ according to the workload balancing heuristic. Lines 14-18 check if the current pair of vehicles leads to a cheaper insertion cost (strict improvement). If the cost current choice of vehicles $\bar{c}$ is very close to $c^*$, then the algorithm still updates the choice of best candidate vehicles $v_1^*$ and $v_2^*$ if the value of the allele associated with the task is $\geqslant 0.5$. This allows the genetic algorithm to reach elements of the solution space otherwise inaccessible. After deciding the best pair of candidates, the algorithm updates the solution by inserting the patient into the routes of the selected caregivers (line 20). Line 21 updates the cost of $s$ using a similar algorithm to \texttt{FindInsertionCost}. Lines 22--25 update the workload of the selected vehicles. Finally, the decoder returns the final solution cost at line 27. Considering $n=|\mathcal{V}|$ and $m=|\mathcal{C}|$, the decoder algorithm has the worst-case complexity of $O(mn^2)$.
\normalcolor

\begin{algorithm}[h]
   \color{black}
   \footnotesize
   \KwIn{A chromosome $\mathit{ch}$.}
   \KwOut{Solution cost.}
   $s \gets $ \texttt{EmptySolution}()\;
   $\mathit{tasks} \gets $ \texttt{Sort}($\mathcal C$, $ch$) \tcc*{Gets the permutation of patients.}
   $\mathit{convHull} \gets $ \texttt{GetKey}($\mathit{ch}$, $|\mathcal C|+1$) $\geqslant 0.5$ \tcc*{Uses the second to last key.}
   $\mathit{wloadHeur} \gets $ \texttt{GetKey}($\mathit{ch}$, $|\mathcal C|+2$) $\geqslant 0.5$ \tcc*{Uses the last key.}
   $\mathit{wload} \gets$ \texttt{Zeros}($|\mathcal V|$) \tcc*{List of zeros.}
   \For {$i \gets 1$ \KwTo $|\mathit{tasks}|$}{
      $c^* \gets \infty$\;
      $(s_1, s_2) \gets $ \texttt{GetSkills}($\mathit{tasks}[i]$)\;
      \For(\tcc*[f]{All pairs of qualified vehicles.}){$(v_1, v_2) \in \mathtt{AllVehicleCombinations}(s_1, s_2)$} {
         $\bar{c} \gets $ \texttt{FindInsertionCost}($s$, $v_1$, $v_2$, $\mathit{tasks}[i]$, $\mathit{convHull}$)\;
         \If(\tcc*[f]{Workload balancing heuristic.}){$\mathit{wloadHeur} = $ \texttt{true}}{
            $\bar{c} \gets \bar{c} + \mathit{wload}[v_1] + (s_2 \neq \mathtt{null}\, ?\, \mathit{wload}[v_2] : 0)$\;
         }
         \If(\tcc*[f]{Strict/non-strict improvement.}){$\bar{c} < c^*$ \textbf{or} ($\bar{c} = c^*$ \textbf{and} \texttt{GetKey}($\mathit{tasks}[i]$) $\geqslant$ 0.5)} {
            $v_1^* \gets v_1$ \;
            $v_2^* \gets v_2$ \;
            $c^* \gets \bar{c}$\;
         }
      }
      \texttt{UpdateRoutes}($s$, $v_1^*$, $v_2^*$, $\mathit{tasks}[i]$) \tcc*{Updates solution structure.}
      \texttt{UpdateCost}($s$, $v_1^*$, $v_2^*$, $\mathit{tasks}[i]$) \tcc*{Updates incumbent.}
      $\mathit{wload}[v_1^*] \gets \mathit{wload}[v_1^*] + \mathtt{ServiceDuration}(task[i], s_1)$       \tcc*{Updates total workload of $v_1^*$.}
      \If {$s_2 \neq \mathtt{null}$} {
         $\mathit{wload}[v_2^*] \gets \mathit{wload}[v_2^*] + \mathtt{ServiceDuration}(task[i], s_2)$ \tcc*{Updates total workload of $v_2^*$.}
      }
   }
   \textbf{return} \texttt{GetCost}($s$)\;
   \caption{Greedy cheapest-insertion constructive heuristic.}
   \label{algorithm:greedy-constructive}
\end{algorithm}

\color{black}
In addition to this description of the decoding algorithm, a few points still need to be explained. First is that function \texttt{GetSkills} returns all the service types requested by the patient and, in case of a single service request, $s_2$ will have the special value \texttt{null}. \texttt{AllVehicleCombinations} is also aware of this behavior, so in such cases, all tuples from the returned list of vehicles will have $v_2$ equals to \texttt{null}. The value of $s_2$ is also used to decide which vehicles are considered when calculating the workload balancing (line 12) and when updating the workload of the vehicles selected to service to a patient (lines 23--25). Note that function \texttt{FindInsertionCost} can also handle the cases in which $v_2$ is \texttt{null}.
\normalcolor

The cost calculation in the insertion heuristic is not straightforward due to the double services and soft time windows. Algorithm~\ref{algorithm:fic} details how this can be implemented, considering all the soft constraints of the problem.
\color{black}
The function starts by querying the current location of vehicles in solution $s$ (line 2). Line 3 calculates the arrival time of $v_1$ at the patient $i$, taking into account the travel time between the current vehicle location and $i$. In case of a single service patient, the algorithm proceeds with lines 5--10, calculating visit tardiness. In case of the \textit{convex hull} strategy to be set on, the algorithm takes a few more steps to subtract the travel times from $\mathit{vpos}_1$ to depot, to then add the travel times from node $i$ to depot (lines 8--10). In the case of a double service patient, the algorithm calculates the arrival time for $v_2$, as  well as the increase in travel times. Lines 14--17 apply similar calculations as lines 8--10. In the case of simultaneous double services, the algorithm calculates the service start times according to line 19. The next two lines calculate the visit tardiness. For a double services patient with precedence, the algorithm calculates the visit times according to the minimum and maximum separation times for patient $i$ (lines 23--27). If the maximum separation time is violated, then a delay is inserted into the start time of the first vehicle (lines 25--27). Violations of the soft time windows are computed in lines 6, 20, and 28, and the maximum tardiness in lines 7, 21, and 29. Finally, the function takes the updated solution quality indicators and returns its cost (lines 32--35). Assuming that all operations inside \texttt{FindInsertionCost} run at constant time, the function has a complexity of $O(1)$.
\normalcolor

\begin{algorithm}[h]
   \color{black}
   \footnotesize
   \KwIn{$s$, $v_1$, $v_2$, $\mathit{task}$, $\mathit{convHull}$.}
   \KwOut{Updated solution cost of appending $\mathit{task}$ to the routes of $v_1$ and $v_2$.}
   $i \gets $ \texttt{GetNode}($\mathit{task}$)\;
   $(\mathit{vpos}_1, \mathit{vpos}_2) \gets \mathtt{VehiclesLocation}(s, v_1, v_2)$ \tcc*{Current location of $v_1$ and $v_2$.}

   $\mathit{arrival}_1 \gets \mathtt{max}\{e_i, \mathtt{LeaveTime}(s, v_1) + \mathtt{TravelTimes}(\mathit{vpos}_1, i)\}$ \;
   $\mathit{dist} \gets \mathtt{TravelTimes}(\mathit{vpos}_1, i)$\;

   \eIf(\tcc*[f]{Processes a single service patient.}){$i \in \mathcal C^\mathrm{s}$}{
      $\mathit{tard} \gets \mathtt{max}\{0, \mathit{arrival}_1 - l_i\}$ \tcc*{Vehicle starts processing immediatelly at $\mathit{arrival}_1$.}
      $\mathit{tmax} \gets \mathit{tard}$\;
      \If{$\mathit{convHull} = \mathtt{true}$}{
         $\mathit{dist} \gets \mathit{dist} - \mathtt{TravelTimes}(\mathit{vpos}_1, 0) + \mathtt{TravelTimes}(i, 0)$\;
      }
   }{
      $\mathit{arrival}_2 \gets \mathtt{max}\{e_i, \mathtt{LeaveTime}(s, v_2) + \mathtt{TravelTimes}(\mathit{vpos}_2, i)\}$\;
      $\mathit{dist} \gets \mathit{dist} + \mathtt{TravelTimes}(\mathit{vpos}_2, i)$\;
      \If{$\mathit{convHull} = \mathtt{true}$}{
         $\mathit{dist} \gets \mathit{dist} - \mathtt{TravelTimes}(\mathit{vpos}_1, 0) + \mathtt{TravelTimes}(i, 0)$\;
         $\mathit{dist} \gets \mathit{dist} - \mathtt{TravelTimes}(\mathit{vpos}_2, 0) + \mathtt{TravelTimes}(i, 0)$\;
      }
      \eIf(\tcc*[f]{Simultaneous double service.}){$\delta^\mathrm{min}_i = \delta^\mathrm{max}_i = 0$} {
         $\mathit{svcStart} \gets \mathtt{max}(\mathit{arrival}_1, \mathit{arrival}_2)$\;
         $\mathit{tard} \gets 2 \cdot \mathtt{max}\{0, \mathit{svcStart} - l_i\}$\;
         $\mathit{tmax} \gets \mathtt{max}\{0, \mathit{svcStart} - l_i\}$\;
      }(\tcc*[f]{Double service with precedence.}){
         $\mathit{svcStart}_1 \gets \mathit{arrival}_1$\;
         $\mathit{svcStart}_2 \gets \mathtt{max}\{\mathit{arrival}_1 + \delta^\mathrm{min}_i, \mathit{arrival}_2\} $\;
         \If(\tcc*[f]{$v_2$ arrived too late at the patient.}){$\mathit{svcStart}_2 - \mathit{svcStart}_1 > \delta^\mathrm{max}_i$}{
            $\mathit{svcStart}_1 \gets \mathit{svcStart}_2 - \delta^\mathrm{max}_i $ \tcc*{Adds some waiting to $v_1$.}
         }
         $\mathit{tard} \gets \mathtt{max}\{0, \mathit{svcStart}_1 - l_i\} + \mathtt{max}\{0, \mathit{svcStart}_2 - l_i\}$\;
         $\mathit{tmax} \gets \mathtt{max}\{\mathtt{max}\{0, \mathit{svcStart}_1 - l_i\}, \mathtt{max}\{0, \mathit{svcStart}_2 - l_i\}\}$\;
      }
   }
   $\mathit{dist} \gets \mathit{dist}~ + $ \texttt{GetTravelTimes}($s$) \;
   $\mathit{tard} \gets \mathit{tard}~ + $ \texttt{GetTotalTardiness}($s$) \;
   $\mathit{tmax} \gets \mathtt{max}\{ \mathit{tmax}, \mathtt{GetTMax}(s)\}$\;
   \textbf{return} $\lambda_1 \cdot \mathit{dist} + \lambda_2 \cdot \mathit{tard} + \lambda_3 \cdot \mathit{tmax}$\;
   \caption{Implementation of \texttt{FindInsertionCost}.}
   \label{algorithm:fic}
\end{algorithm}

\color{black}
Despite its similarity to the decoding algorithm proposed by \citet{kummer2020}, our new decoder has some important additions that allow the genetic algorithm to reach a larger amount of the solution space of the HHCRSP. These additions--notably the \textit{convex hull} strategy--allow our BRKGA to achieve solutions otherwise inaccessible by previous work in the literature. Similarly, the \textit{workload balancing heuristic} can also help to improve solution quality when a problem instance has similar vehicles with respect to qualification levels, which indirectly guide to solutions with improved travel times.
\normalcolor

\section{Computational results}
\label{section:computational results}

This section presents computational results for our proposed metaheuristic. We describe the benchmark dataset used in our computational experiments and the benchmark machine. We also compare our results with all the previous publications that use this same dataset, and we highlight which solution methods work best for each instance tested.

\subsection{Computational environment and benchmark dataset}
\label{subsection:bench-computer}

We use in our experiments an Intel Xeon E5-2697 v2 computer running at 2.70 GHz, with 64 GB of memory. Our code is implemented in C++, and we compile our binary using the GNU G++ compiler, version 7.5.0 running Ubuntu 18.04 with Linux kernel 4.15.0.
We use the open-source \href{https://github.com/ceandrade/brkga_mp_ipr_cpp}{\texttt{brkga\_mp\_ipr\_cpp}} framework of \citet{andrade2021}. This framework not only allows developing standard BRKGA algorithms, including the variant with an island model, but it also supports the IPR and the MP mating described in the former sections. This way, most of our implementation effort is to develop a problem-specific decoder compatible with the \citeauthor{andrade2021} framework, altogether with some auxiliary code to handle the problem data.
The \texttt{brkga\_mp\_ipr\_cpp} framework supports the new C++17 standard. Similar to its predecessor, the \texttt{brkgaAPI} of \citet{toso2015}, it also supports multi-core processing through OpenMP directives. Thus, we take advantage of our benchmark machine's parallel capabilities to decode individuals in parallel using four \textit{real} cores--we explicitly disabled the \textit{hyperthreading} feature during our experiments.

We test out metaheuristic algorithm using the benchmark dataset proposed by \citet{mankowska2014}. This dataset is comprised of seven instance subsets, each one composed of ten synthetic instances. \textcolor{black}{Table~\ref{table:instances-families} highlights the general properties of each subset.} The columns identify the instance family, the total number of patients, and the number of single service and double service patients, respectively. The last column presents the number of caregivers. The set of service types $\mathcal{S} = \{1,2,3,4,5,6\}$ is the same for all the 70 instances, and the selection of the requested service types \textcolor{black}{per patient is at} random. Each instance comprises 30\% of double service patients, 15\% requiring simultaneous double services, and 15\% requiring \textcolor{black}{double services with} precedence constraints. The other 70\% are single service patients. Note that single service patients require just one service type, and double service patients require exactly two. The caregiver skills are also set at random. Half of the caregivers can perform up to three service types from the set $\{1,2,3\}$. The same happens to the other half of $\mathcal{V}$, considering the service types $\{4,5,6\}$. The authors guarantee at least one caregiver capable of performing each service type. All the points (the depot and patient homes) were generated within a 100$\times$100 grid, and the travel times are symmetric. Each patient has a two-hour time-windows drawn from a time horizon of ten hours.

\bgroup
\begin{table}[!htb]
	\caption{Characteristics of the benchmarking dataset proposed by \cite{mankowska2014}. \textcolor{black}{For each instance subset, we indicate the total number of patients $|\mathcal C|$, and the number of single $|\mathcal C^\mathrm{s}|$ and double service patients $|\mathcal C^\mathrm{d}|$, respectively. Column $|\mathcal V|$ indicates the number of caregivers considered}.}
	\label{table:instances-families}
	\centering
	\footnotesize
	\begin{tabular}{lrrrr}
      \toprule
      Instance subset & $|\mathcal{C}|$ & $|\mathcal{C}^\mathrm{s}|$ & $|\mathcal{C}^\mathrm{d}|$& $|\mathcal{V}|$\\
      \midrule
      A & 10  & 7   &   3 & 3 \\
      B & 25  & 17  &   8 & 5 \\
      C & 50  & 35  &  15 & 10 \\
      D & 75  & 52  &  23 & 15 \\
      E & 100 & 70  &  30 & 20 \\
      F & 200 & 140 &  60 & 30 \\
      G & 300 & 200 & 100 & 40 \\
      \midrule
      \color{black}
      \# of instances & \multicolumn{4}{c}{\color{black} 70}\\
      \bottomrule
   \end{tabular}
\end{table}
\egroup

Lastly, we use the weights $\lambda_1 = \lambda_2 = \lambda_3 =\;\!\! ^1\!/_3$ in the objective function. These are the same weights used in all previous publications that use the \citet{mankowska2014} dataset.

\subsection{Improved lower bounds with CPLEX 20.1.0.0}
\label{subsection:new-best-lb}

During computational experiments with CPLEX 12.3, \citet{mankowska2014} attempted to solve their instance dataset by applying the MIP model (\ref{model_obj}--\ref{model_dtt_dom}). According to the authors, only instances with up to ten patients could be solved to optimality, and feasible solutions could only be found for other test cases with 25 patients. For these experiments with the MIP, the authors also report the best lower bound found within 10 hours of solver runtime. Thanks to the generation of additional cuts, its value improves during the branch-and-cut phase due. Thus, these best lower bounds are stronger than the simple lower bounds from the MIP linear relaxation. For the test cases with 200 and 300 patients, the authors simplify the experiments by relaxing the \textcolor{black}{route inter-dependency} constraints (\ref{model_sync1}) and (\ref{model_sync2}). This way, the best lower bound provided for these instances are (provably) weaker than the best lower bounds for the HHCRSP.

Prior to assessing the solution quality for our proposed metaheuristic algorithm, we first experiment with the \citeauthor{mankowska2014} model to obtain stronger new best lower bounds (LB$^+$), using the most updated version available of the CPLEX solver. In our experiment, we consider all the constraints for the MIP (\ref{model_obj}--\ref{model_dtt_dom}). We also use constraints (\ref{model_x_dom}) to skip generating decision variables that are trivially fixed to zero due to either the lack of caregiver skills or missing patient request for some service types. Such preprocessing also allows us to skip generating empty flow constraints (\ref{model_flow_conservation}). With these minor improvements, we significantly reduce the MIP memory consumption by a third for instances with 300 patients. \textcolor{black}{We solve the MIP using CPLEX 20.1.0.0 and we applied three configurations to the solver, aiming to test the standard optimizer performance (Defaults), and to test the effectiveness of our strategies for reducing the growth in memory consumption as the search progresses (configurations \#1 and \#2).} We achieve the following parameter setting presented in Table \ref{table:cplex-param-best-lb} by inspecting the optimizer documentation \citep{cplex201-documentation}. \textcolor{black}{Reducing the number of threads available to CPLEX reduces the growth in memory consumption as the branch-and-cut algorithm proceeds. The solver also has a ``memory emphasis'' flag that disables some memory-intensive pre-processing routines, at expanse of worsening the solver performance.}

\bgroup
\begin{table}[!htb]
   \color{black}
   \centering
   \caption{Parameter setting for CPLEX while obtaining strong best lower bounds.}
   \label{table:cplex-param-best-lb}
   \begin{tabular}{llll}
      \toprule
      Parameter & Defaults & Configuration \#1 & Configuration \#2\\
      \midrule
      \# threads & (automatic) & 1 & 1\\
      Memory emphasis & off & on & on\\
      MIP warmstart & off & off & on\\
      \bottomrule
   \end{tabular}
\end{table}
\egroup

In addition to the tweaked parameters, \textcolor{black}{preliminary experiments indicate} that CPLEX can greatly benefit from a \textit{warm start} from a feasible solution to enhance further both memory usage and to produce better values of LB$^+$. This way, we use the genetic algorithm of \citet{kummer2020} to produce such initial solutions. We use most of the parameter settings proposed in their paper, but we limit the number of evolved generations to 600 to reduce the time spent on the warm start. In these experiments, we use the same computer we describe in Section \ref{subsection:bench-computer}. We use CPLEX command line interface to run our tests\textcolor{black}{, and we} set a time limit of two hours for each run of the solver. \textcolor{black}{To better represent a \textit{standard personal computer}, we also limited CPLEX to use up to 8 GB of memory.}

Table~\ref{table:new-best-lb-cplex201-short} presents our new average best lower bounds for each instance subset. The first column presents the instance subset's name, followed by the computational time and the best lower bounds reported by \citet{mankowska2014}. \textcolor{black}{Note that we adjusted the times reported by \citeauthor{mankowska2014}, as explained in Subsection \ref{subsection:comparison-lit-ls}. The next columns present our average results for each instance subset, for the standard CPLEX configuration, and configurations \#1 and \#2. For each configuration, we report the average solver runtime in seconds, the average number of branch-and-cut nodes explored, and the average LB$^+$. At the bottom of the table, we present the number of instances in which each experiment produced the best lower bound (\#~Best), the number of proved optimal solutions (\#~Opt), the number of instances that achieved a time limit of 2 hours (\#~TL), and the number of runs that were stopped due to out-of-memory (\#~OOM). }

\bgroup
\begin{table}[!htb]
   \color{black}
   \caption{New best lower-bound found with CPLEX 20.1.0.0. Values in bold indicate the best lower-bound available.}
   \label{table:new-best-lb-cplex201-short}
   \scriptsize
   \centering
   \begin{tabular}{lrrrrrrrrrrr}
      \toprule
      \multirow{2}[2]{*}{\makecell{Inst.\\subset}} &
      \multicolumn{2}{c}{Literature} &
      \multicolumn{3}{c}{Defaults} &
      \multicolumn{3}{c}{Configuration \#1} &
      \multicolumn{3}{c}{Configuration \#2}\\

      \cmidrule(l){2-3}
      \cmidrule(l){4-6}
      \cmidrule(l){7-9}
      \cmidrule(l){10-12}

      & T (sec.) & LB$^+$
      & T (sec.) & \# Nodes & LB$^+$
      & T (sec.) & \# Nodes & LB$^+$
      & T (sec.) & \# Nodes & LB$^+$\\
      \midrule

      A & 5.47 & \textbf{225.18} & 0.26 & 2158.70 & \textbf{225.18} & 0.61 & 1038.20 & \textbf{225.18} & 0.27 & 656.70 & \textbf{225.18} \\
      B & 37,908 & 343.11 & 3169.94 & 2,445,294.20 & 387.14 & 7200.00 & 1,867,494.20 & 345.90 & 3250.88 & 705,862.00 & \textbf{391.29} \\
      C & 37,908 & 342.62 & 4640.88 & 357,600.10 & 384.24 & 7200.00 & 217,070.70 & 382.00 & 7200.01 & 222,574.10 & \textbf{393.75} \\
      D & 37,908 & 377.36 & 5595.87 & 123,144.00 & 399.73 & 7200.01 & 32,363.10 & 408.71 & 7200.02 & 38,002.30 & \textbf{412.71} \\
      E & 37,908 & 404.53 & 4952.80 & 58,249.20 & 402.10 & 7200.02 & 9,987.50 & \textbf{428.47} & 7200.06 & 12,679.80 & 425.55 \\
      F & 37,908 & 435.26 & 3210.13 & 2696.30 & 492.88 & 7200.37 & 1,682.90 & 530.97 & 7200.40 & 690.90 & \textbf{536.43} \\
      G & 37,908 & 462.12 & -- & -- & -- & 7200.84 & 358.00 & \textbf{620.76} & 7200.62 & 336.40 & 604.59 \\
      \midrule
      \multicolumn{1}{c}{\# Best} & \multicolumn{2}{c}{9} & \multicolumn{3}{c}{31} & \multicolumn{3}{c}{29}   & \multicolumn{3}{c}{46}\\
      \multicolumn{1}{c}{\# Opt} & \multicolumn{2}{c}{10} & \multicolumn{3}{c}{17} & \multicolumn{3}{c}{10}   & \multicolumn{3}{c}{17}\\
      \multicolumn{1}{c}{\# TL} & \multicolumn{2}{c}{60} & \multicolumn{3}{c}{17} & \multicolumn{3}{c}{60}   & \multicolumn{3}{c}{53}\\
      \multicolumn{1}{c}{\# OOM} & \multicolumn{2}{c}{0} & \multicolumn{3}{c}{36} & \multicolumn{3}{c}{0}   & \multicolumn{3}{c}{0}\\
      \bottomrule
   \end{tabular}
\end{table}
\egroup

On average, the newer version of CPLEX can find better LB$^+$ values than the previously reported in the literature, although the most remarkable difference is that we use only 20\% of the processing time of \citet{mankowska2014} to produce these new results.
We also managed to prove the optimal solutions to \textcolor{black}{7 out of 10 instances} of subset B, which justifies the average running time of \textcolor{black}{3169.94 seconds for this group of instances (using CPLEX defaults).} These results are obtained at the expense of exploring gigantic branch-and-cut trees with more than $10^6$ nodes on average. Despite that, we observed no trees consuming more than 500 MB of memory for subset B. Naturally, the number of nodes explored drops as the problem size increases, as a consequence of the large time spent solving LP subproblems. The smaller improvement of LB$^+$ on subsets D and E is related to some instances that are particularly difficult to solve, e.g., E3, for which the LB$^+$ improvement is only \textcolor{black}{3.90\% between the literature and configuration~\#1, and 3.27\% between the literature and configuration~\#2.} We believe this issue may be related to phase transition between easy and hard instances of the HHCRSP, which is often discussed in the literature of other optimization problems.

\subsection{Automatic algorithm configuration with \textit{irace}}
\label{subsection:irace}

\textcolor{black}{As the reader may expect}, the presence of several components makes our proposed genetic algorithm heavily parameterized. \textcolor{black}{In such scenarios, a state-of-art approach to configure an algorithm is to apply } an automatic algorithm configuration (AAC) tool \textcolor{black}{to automate the finding of good set of parameters.} The AAC approach has several inherent advantages over the manual configuration procedure. It reduces the human effort, allowing the practical evaluation of thousands of configurations automatically. It also helps prevent experimental errors or \textcolor{black}{preconceptions} in consequence of any human bias with respect to algorithm behavior \citep{eggensperger2019-aac-best-practices}. Tools like \textit{irace} \citep{lopez-ibanes2016-irace} require a very minimal effort to set up an AAC experiment. First, we need to specify the \textit{range of values} for each parameter we want \textit{irace} to configure. We also need to provide a set of \textit{training instances} and a \textit{budget} of maximum runs of the algorithm to tune. \textit{Irace} then evaluates many configurations automatically by iteratively sampling new ones, evaluating them over the training dataset, and then refining the sampling model towards promising configurations generated so far.

\textcolor{black}{To configure our metaheuristic algorithm, we employed \textit{irace} version 3.4.1 (R 3.4.4), and we used the same machine we described in Section~\ref{subsection:bench-computer}. Regarding metaheuristic components, we set three parameters for controlling the ``traditional'' BRKGA elements; three parameters for the multi-parent mating component; three parameters controlling the island model; and four parameters controlling the implicit path relinking. We consider a total of \textcolor{black}{13} parameters listed in Table \ref{table:irace-params}. The first column presents the parameter description, and the second column presents the domain of each parameter. The third column depicts the range of values we allow \textit{irace} to evaluate. The last column presents the parameter setting set by \textit{irace} as the best configuration found after a budget of 300 runs of our BRKGA-based algorithm. Additionally, we use the G instance subset as training instances during the AAC experiment.}

\begin{table}[!htb]
   \color{black}
   \centering
   \footnotesize
   \caption{Parameters and range of values configured by \textit{irace}.}
   \label{table:irace-params}
   \begin{tabular}{llrr}
      \toprule
      Parameter description & Type & Range of values & Best config.\\
      \midrule

      Population size & \texttt{int} & $\left[500, 1500\right]$ & 1462\\

      Elite (\%) & \texttt{real} & $\left[0.05, 0.4\right]$ & 0.30678\\

      Mutant (\%) & \texttt{real} & $\left[0.01, 0.3\right]$ & 0.07575\\

      \# of parents & \texttt{int} & $\left[2, 6\right]$ & 5\\
      \# of elite parents & \texttt{int} & $\left[1, 5\right]$ & 4\\

      Bias function & \texttt{categorical} &
      \makecell[r]{
         \{\texttt{constant}, \texttt{cubic},\\
         \texttt{exponential}, \texttt{linear}, \\
         \texttt{loginverse}, \texttt{quadatic}\}
      } & \texttt{constant}\\

      \midrule

      Independent populations & \texttt{int} & $\left[2, 6\right]$ & 2\\

      Immigrants & \texttt{int} & $\left[3, 150\right]$ & 73\\

      Exchange elite frequency & \texttt{int} & $\left[30, 180\right]$ & 167\\

      \midrule

      IPR pairs & \texttt{int} & $\left[1, 100\right]$ & 95\\

      IPR selection criteria & \texttt{categorical} & \{\texttt{best}, \texttt{random}\} & \texttt{random}\\

      IPR path (\%) to explore & \texttt{real} & $\left[0.01, 1.0\right]$ & 0.33754\\


      IPR frequency & \texttt{int} & $\left[5, 100\right]$ & 40\\

      \bottomrule
   \end{tabular}
\end{table}

\color{black}
Some of the parameters required by our algorithm were excluded from the AAC experiment, and we set them to fixed values. For example, \texttt{brkga\_mp\_ipr\_cpp} supports both \textit{direct} and \textit{permutation} implicit path-relinking, but we fixed our implementation to always the permutation IPR due to the properties mentioned by \citet{andrade2021}. We also set the minimum distance threshold for IPR runs to zero,  leaving to \textit{irace} to decide the frequency which IPR should be applied, as well as how many pairs of individuals to try.
\normalcolor



From \textcolor{black}{our} computational experiments, we observe the island model helps to cope with the typical issue diversity loss among the evolutionary process. Using the parameter setting from Table~\ref{table:irace-params}, 16\% of the elite sets immigrate from one island to the others periodically after 167 generations, keeping some diversity as the search progresses. Additionally, the implicit path-relinking also improves the diversity by generating hybrid solution from individual of distinct islands. Typically these intermediate individuals improve the diversity of both elite sets over the time. \textcolor{black}{Considering the standard deviation metric applied to individuals' fitness,} we observe an increase in the diversity of the elite set each time a new solution is found, mainly by the mating process, by immigration, and by the IPR, in this order.

\color{black}
The stopping criterion of our genetic algorithm is calculated dynamically, and it is proportional to the number of patients of each test cases. This way, our algorithm stops the search after achieving $\frac{|\mathcal C|}{2}$ generations without improvement.
\normalcolor

\subsection{Comparison with local-search-based method from literature}
\label{subsection:comparison-lit-ls}

In this section, we compare our results with those of local-search-based methods from the literature. As we mention previously, several authors test their methods against the same dataset of \citet{mankowska2014}, enabling us to compare distinct approaches with our proposal. Despite that, we did not replicated all the experiments from the literature. Instead, we compare against the results from their original publications. We then correct the published processing times to our benchmark machine using the \textit{PassMark} scores as comparison criteria \citep{passmark-documentation}. Table~\ref{table:speed-factors} present these factors for \citet{mankowska2014} and  \citet{lasfargeas2019}. The first column presents the reference, and the second column presents the description of the machine used in their experiments. The third column presents the machine score according to \textit{PassMark}. The last column presents a speed factor we use to correct their published processing times. As the table shows, our machine is actually \textit{slower} than the ones used in all previous works, so we might expect to run our metaheuristic algorithm \textit{faster} in their machines than ours. \textcolor{black}{Furthermore, the source code of both \citet{neto2019} and \citet{kummer2020} are available online, so we could re-run their experiments in our benchmark machine.}

\begin{table}[htb]
	\caption{
		Comparison of computational environments. Neither \citet{mankowska2014} nor \cite{lasfargeas2019} specify which machine they use in their experiments. Considering the date of the publications and the documentation available on the Intel website, we guess that \citeauthor{mankowska2014} use an Intel i3-4130, and that \citeauthor{lasfargeas2019} use an Intel i7-6700K \textcolor{black}{as} their benchmark machines.
	}
	\label{table:speed-factors}
	\centering
	\footnotesize
	\begin{tabular}{llrr}
		\toprule
		Publication & Machine description & PassMark score & Factor\\
		\midrule
		\citet{mankowska2014} & Intel Core at 3.40 GHz & 1887 & 1.0530\\
		\citet{lasfargeas2019} & Intel i7 at 4.00 GHz & 2523 & 1.4080\\
		This work & Intel Xeon E5-2697 v2 at 2.70 GHz & 1792 & 1.0000\\
		\bottomrule
	\end{tabular}
\end{table}

Table~\ref{table:comparison-literature-others} presents the previous results from the literature, together with the results for the BRKGA of this paper. \textcolor{black}{We solve each instance from \citeauthor{mankowska2014} dataset 20 times, using the numbers between 1 to 20 as seeds. We also rerun the fix-and-optimize \textit{matheuristic} of \citet{neto2019} 20 times, using the same seeds as to our algorithm.} The first column presents the instance name. The second column presents the new best lower bounds we discussed in Section \ref{subsection:new-best-lb}.
The next two columns present the results from \citet{mankowska2014}, respectively, \textcolor{black}{the solve times adjusted using the factors we discussed earlier, and the solution cost reported by the authors. Note that the algorithms proposed by \citet{mankowska2014} are all deterministic, so the authors reported results for a single run on each instance. The next four columns present the results from \citet{lasfargeas2019}: their solve times--also corrected using the speed factors, the best solution value they produced, and the average and standard deviation of the solution cost. According to their report, each instance was solved 40 times. The next four columns present the data as to \citet{lasfargeas2019}, but these columns report the values for the fix-and-optimize \textit{matheuristic} proposed by \citet{neto2019}. Again, the last four columns present these same data, but for the BRKGA-based algorithm proposed in this work. At the bottom of the table, we present the average solve time for each approach (Avg time (sec.)), the number of instances which each method produced the best solution value (\#~Best), and the number of times which each method produced the best average solution value (\#~Best~avg). To ease our writing, we refer to \citet{mankowska2014}, \citet{lasfargeas2019}, and \citet{neto2019} by MBB2014, LGS2019, and KBA2019, respectively.}


\newpage
\bgroup
\scriptsize
\renewcommand{\tabcolsep}{3pt}
\renewcommand{\arraystretch}{1.05}
\color{black}
\begin{longtable}{lrrrrrrrrrrrrrrr}
   \caption{Extensive computational results comparing the performance of local search-based method from the literature, and our proposed BRKGA.}
   \label{table:comparison-literature-others}\\
   \toprule
   \multirow{2}[5]{*}{Inst.} &
   \multirow{2}[5]{*}{LB$^+$} &
   \multicolumn{2}{c}{MMB2014} &
   \multicolumn{4}{c}{LGS2019} &
   \multicolumn{4}{c}{KBA2019} &
   \multicolumn{4}{c}{BRKGA}\\
   \cmidrule(r){3-4}
   \cmidrule(r){5-8}
   \cmidrule(r){9-12}
   \cmidrule(){13-16}
   & &
   \makecell{Time\\(sec)} & Cost &
   \makecell{Time\\(sec)} & Best & Avg & SD &
   \makecell{Time\\(sec)} & Best & Avg & SD &
   \makecell{Time\\(sec)} & Best & Avg & SD \\
   \midrule
   \endfirsthead

   \caption*{Table \ref{table:comparison-literature-others}: (Continued) Extensive computational results comparing the performance of local search-based method from the literature, and our proposed BRKGA.}\\
   \toprule
   \multirow{2}[5]{*}{Inst.} &
   \multirow{2}[5]{*}{LB$^+$} &
   \multicolumn{2}{c}{MMB2014} &
   \multicolumn{4}{c}{LGS2019} &
   \multicolumn{4}{c}{KBA2019} &
   \multicolumn{4}{c}{BRKGA}\\
   \cmidrule(r){3-4}
   \cmidrule(r){5-8}
   \cmidrule(r){9-12}
   \cmidrule(){13-16}
   & &
   \makecell{Time\\(sec)} & Cost &
   \makecell{Time\\(sec)} & Best & Avg & SD &
   \makecell{Time\\(sec)} & Best & Avg & SD &
   \makecell{Time\\(sec)} & Best & Avg & SD \\
   \midrule
   \endhead

   \midrule
   \endfoot

   \bottomrule
   \endlastfoot

   A1 & 218.20 & 2.11 & \textbf{218.20} & <1 & \textbf{218.20} & 224.30 & 14.60 & 0.15 & \textbf{218.20} & 219.19 & 4.44 & 0.20 & \textbf{218.20} & \textbf{218.20} & 0.00 \\
   A2 & 246.63 & 5.27 & \textbf{246.60} & <1 & \textbf{246.60} & 258.10 & 31.60 & 0.15 & \textbf{246.63} & \textbf{246.63} & 0.00 & 0.17 & \textbf{246.63} & \textbf{246.63} & 0.00 \\
   A3 & 305.86 & 7.37 & \textbf{305.90} & <1 & \textbf{305.90} & 358.80 & 38.50 & 0.39 & \textbf{305.86} & \textbf{305.86} & 0.00 & 0.16 & \textbf{305.86} & \textbf{305.86} & 0.00 \\
   A4 & 186.90 & 8.42 & \textbf{186.90} & <1 & \textbf{186.90} & 196.40 & 11.10 & 1.10 & \textbf{186.90} & 189.25 & 7.24 & 0.16 & \textbf{186.90} & \textbf{186.90} & 0.00 \\
   A5 & 189.54 & 2.11 & \textbf{189.50} & <1 & \textbf{189.50} & 216.50 & 36.60 & 0.19 & \textbf{189.54} & 194.00 & 13.73 & 0.18 & \textbf{189.54} & \textbf{189.55} & 0.01 \\
   A6 & 200.10 & 2.11 & \textbf{200.10} & <1 & \textbf{200.10} & \textbf{200.10} & 0.00 & 0.15 & \textbf{200.10} & \textbf{200.11} & 0.03 & 0.15 & \textbf{200.13} & \textbf{200.13} & 0.01 \\
   A7 & 225.40 & 1.05 & \textbf{225.40} & <1 & \textbf{225.40} & 232.40 & 24.20 & 0.11 & \textbf{225.37} & \textbf{225.37} & 0.00 & 0.18 & \textbf{225.37} & \textbf{225.37} & 0.00 \\
   A8 & 232.05 & 4.21 & \textbf{232.00} & <1 & \textbf{232.00} & 281.40 & 47.60 & 0.16 & \textbf{232.05} & \textbf{232.05} & 0.00 & 0.16 & \textbf{232.05} & \textbf{232.05} & 0.00 \\
   A9 & 222.30 & 21.06 & \textbf{222.30} & <1 & \textbf{222.30} & 225.40 & 4.00 & 2.10 & \textbf{222.30} & \textbf{222.30} & 0.00 & 0.20 & \textbf{222.30} & 223.75 & 3.57 \\
   A10 & 225.01 & 1.05 & \textbf{225.00} & <1 & \textbf{225.00} & \textbf{225.00} & 0.00 & 0.09 & \textbf{225.01} & 226.55 & 6.91 & 0.13 & \textbf{225.01} & \textbf{225.01} & 0.00 \\
   \midrule
   Avg & 225.20 & 5.48 & \textbf{225.19} & <1 & \textbf{225.19} & 241.84 & 20.82 & 0.46 & \textbf{225.19} & 226.13 & 3.23 & 0.17 & \textbf{225.20} & \textbf{225.35} & 0.36 \\
   \midrule	
   B1 & 428.10 & <1 & 458.90 & 74.76 & 434.10 & 552.80 & 93.40 & 36.17 & \textbf{428.10} & 434.79 & 8.06 & 0.78 & \textbf{428.10} & \textbf{428.53} & 0.15 \\
   B2 & 476.05 & 37,908.00 & 476.20 & 39.00 & \textbf{476.00} & 561.30 & 61.40 & 10.85 & \textbf{476.05} & 483.03 & 17.71 & 0.87 & \textbf{476.05} & \textbf{476.92} & 2.34 \\
   B3 & 399.09 & 37,908.00 & 399.20 & 89.41 & \textbf{399.10} & 527.60 & 72.50 & 177.66 & \textbf{399.09} & 419.51 & 12.07 & 0.99 & 402.80 & \textbf{409.29} & 4.78 \\
   B4 & 411.30 & 37,908.00 & 576.00 & 94.05 & 414.00 & 509.70 & 74.50 & 17.48 & \textbf{411.30} & 439.23 & 17.23 & 1.13 & 422.06 & \textbf{430.46} & 8.81 \\
   B5 & 366.34 & 25.76 & 391.10 & 19.29 & 385.60 & 496.90 & 98.10 & 64.06 & \textbf{366.34} & 390.17 & 20.30 & 0.98 & 369.44 & \textbf{375.15} & 3.88 \\
   B6 & 405.58 & 99.09 & 534.70 & 61.53 & 447.80 & 611.80 & 129.90 & 220.15 & \textbf{441.70} & 516.99 & 84.59 & 1.22 & 470.59 & \textbf{470.70} & 0.16 \\
   B7 & 328.67 & 14.81 & 355.50 & 86.59 & \textbf{328.70} & 398.80 & 64.80 & 35.97 & \textbf{328.67} & 334.33 & 9.72 & 0.93 & \textbf{328.67} & \textbf{328.67} & 0.00 \\
   B8 & 357.68 & 37,908.00 & 357.80 & 111.65 & 359.70 & 488.70 & 116.20 & 30.71 & \textbf{357.68} & 373.95 & 16.85 & 0.70 & \textbf{357.68} & \textbf{359.40} & 0.66 \\
   B9 & 330.30 & 37,908.00 & 403.80 & 87.44 & 404.10 & 483.40 & 60.30 & 185.51 & \textbf{402.67} & 410.59 & 18.52 & 0.91 & 404.11 & \textbf{404.29} & 0.23 \\
   B10 & 420.99 & 31.34 & 500.40 & 12.25 & \textbf{462.70} & 616.80 & 147.70 & 184.17 & \textbf{462.75} & 479.48 & 17.31 & 0.85 & 469.58 & \textbf{469.58} & 0.00 \\
   \midrule
   Avg & 392.41 & 21079.00 & 445.36 & 67.60 & 411.18 & 524.78 & 91.88 & 96.27 & \textbf{407.43} & 428.21 & 22.24 & 0.93 & 412.91 & \textbf{415.30} & 2.10 \\
   \midrule
   C1 & 459.25 & 180.13 & 1123.60 & 135.45 & 974.20 & 1350.40 & 365.30 & 1058.09 & \textbf{957.05} & 1001.48 & 60.67 & 3.09 & 969.11 & \textbf{973.87} & 3.52 \\
   C2 & 373.94 & 114.99 & 673.80 & 149.81 & 605.10 & 685.50 & 55.60 & 330.47 & \textbf{582.99} & 610.95 & 43.52 & 2.88 & 584.18 & \textbf{587.00} & 3.26 \\
   C3 & 390.48 & 98.58 & 642.40 & 154.60 & 562.90 & 698.20 & 82.70 & 453.59 & 558.75 & 644.42 & 91.42 & 2.86 & \textbf{549.63} & \textbf{552.52} & 3.96 \\
   C4 & 371.99 & 76.20 & 580.40 & 158.26 & 521.90 & 630.40 & 101.80 & 180.66 & \textbf{507.38} & 527.13 & 25.72 & 2.97 & 520.13 & \textbf{524.15} & 2.88 \\
   C5 & 464.97 & 86.69 & 754.60 & 161.78 & 683.10 & 822.60 & 119.30 & 316.68 & \textbf{667.53} & 687.64 & 20.16 & 3.35 & 668.65 & \textbf{685.92} & 13.92 \\
   C6 & 360.73 & 208.76 & 951.60 & 163.19 & 854.60 & 1010.60 & 146.40 & 989.14 & \textbf{822.85} & 900.74 & 82.23 & 2.92 & 841.48 & \textbf{846.83} & 2.40 \\
   C7 & 354.15 & 82.90 & 577.40 & 154.04 & 529.20 & 572.50 & 29.70 & 351.53 & \textbf{521.89} & \textbf{540.31} & 10.63 & 3.37 & 533.92 & 541.88 & 5.94 \\
   C8 & 375.52 & 60.75 & 540.60 & 156.01 & \textbf{471.00} & 522.80 & 29.80 & 115.40 & 476.66 & 489.45 & 10.30 & 3.46 & 475.96 & \textbf{478.39} & 2.84 \\
   C9 & 355.29 & 98.60 & 608.70 & 162.48 & 551.10 & 642.70 & 77.60 & 346.79 & \textbf{535.87} & 572.36 & 29.68 & 3.44 & 545.18 & \textbf{558.54} & 15.13 \\
   C10 & 431.18 & 75.80 & 679.30 & 139.39 & 608.90 & 653.00 & 35.60 & 341.12 & \textbf{590.26} & 617.35 & 26.91 & 2.73 & 611.03 & \textbf{614.59} & 3.74 \\
   \midrule
   Avg & 393.75 & 108.34 & 713.24 & 153.50 & 636.20 & 758.87 & 104.38 & 448.35 & \textbf{622.12} & 659.18 & 40.12 & 3.10 & 629.93 & \textbf{636.37} & 5.76 \\
   \midrule
   D1 & 492.09 & 5.27 & 1321.80 & 201.34 & 1278.20 & 1498.80 & 199.00 & 2248.63 & \textbf{1149.70} & \textbf{1202.98} & 66.99 & 7.99 & 1193.21 & 1215.79 & 11.11 \\
   D2 & 384.68 & 4.21 & 892.70 & 237.53 & 746.90 & 914.30 & 97.90 & 1017.40 & 690.21 & 761.25 & 50.31 & 7.24 & \textbf{679.58} & \textbf{695.99} & 11.49 \\
   D3 & 380.05 & 4.21 & 819.40 & 218.80 & 678.60 & 817.80 & 80.90 & 1097.04 & \textbf{643.23} & 680.90 & 36.63 & 8.50 & 644.16 & \textbf{650.22} & 3.65 \\
   D4 & 418.94 & 4.21 & 877.40 & 209.09 & 809.70 & 1073.10 & 197.50 & 1329.47 & 798.99 & 828.36 & 23.85 & 7.22 & \textbf{795.15} & \textbf{827.28} & 12.32 \\
   D5 & 415.81 & 5.27 & 872.10 & 211.62 & 777.00 & 924.90 & 120.80 & 1231.71 & \textbf{688.53} & 741.15 & 35.31 & 7.72 & 693.83 & \textbf{702.68} & 3.91 \\
   D6 & 392.08 & 5.27 & 835.20 & 217.68 & 768.60 & 886.60 & 97.40 & 1011.85 & \textbf{712.15} & 764.51 & 31.68 & 7.91 & 731.71 & \textbf{743.64} & 5.97 \\
   D7 & 372.49 & 6.32 & 706.30 & 236.68 & 600.10 & 680.40 & 31.60 & 1121.08 & 595.22 & 619.66 & 20.47 & 7.83 & \textbf{586.10} & \textbf{597.25} & 5.68 \\
   D8 & 409.35 & 4.21 & 811.40 & 210.92 & 715.50 & 775.80 & 31.10 & 975.76 & 666.09 & 712.81 & 31.97 & 8.10 & \textbf{658.49} & \textbf{669.83} & 7.02 \\
   D9 & 385.89 & 6.32 & 842.70 & 219.65 & 741.00 & 818.20 & 46.50 & 661.42 & \textbf{671.23} & 726.78 & 28.43 & 9.15 & 689.83 & \textbf{710.32} & 10.95 \\
   D10 & 485.63 & 3.16 & 1306.60 & 243.72 & 1424.60 & 1867.70 & 258.60 & 1733.73 & 1239.79 & 1329.02 & 88.15 & 6.87 & \textbf{1189.32} & \textbf{1280.92} & 62.72 \\
   \midrule
   Avg & 413.70 & 4.84 & 928.56 & 220.70 & 854.02 & 1025.76 & 116.13 & 1242.81 & \textbf{785.51} & 836.74 & 41.38 & 7.85 & 786.14 & \textbf{809.39} & 13.48 \\
   \newpage
   E1 & 430.36 & 17.90 & 1604.90 & -- & -- & -- & -- & 4201.41 & 1337.87 & 1466.10 & 121.41 & 17.19 & \textbf{1327.72} & \textbf{1340.36} & 8.44 \\
   E2 & 444.88 & 10.53 & 1101.90 & -- & -- & -- & -- & 3860.38 & 858.38 & 924.91 & 51.62 & 17.14 & \textbf{829.79} & \textbf{865.05} & 45.91 \\
   E3 & 454.27 & 14.74 & 986.40 & -- & -- & -- & -- & 3246.19 & 795.72 & 883.85 & 43.01 & 16.70 & \textbf{789.56} & \textbf{806.53} & 17.56 \\
   E4 & 412.08 & 20.01 & 871.00 & -- & -- & -- & -- & 3710.59 & 732.28 & 791.98 & 40.14 & 14.77 & \textbf{723.87} & \textbf{728.96} & 3.45 \\
   E5 & 416.62 & 20.01 & 1018.00 & -- & -- & -- & -- & 4643.35 & 791.18 & 857.18 & 34.80 & 17.03 & \textbf{780.04} & \textbf{817.13} & 34.10 \\
   E6 & 416.60 & 20.01 & 1003.00 & -- & -- & -- & -- & 4216.42 & 831.76 & 881.96 & 39.75 & 18.28 & \textbf{779.82} & \textbf{793.68} & 8.15 \\
   E7 & 389.57 & 21.06 & 921.10 & -- & -- & -- & -- & 3565.23 & 744.15 & 813.42 & 42.63 & 18.02 & \textbf{705.79} & \textbf{715.46} & 7.89 \\
   E8 & 433.89 & 20.01 & 884.60 & -- & -- & -- & -- & 3889.80 & 745.18 & 808.44 & 30.05 & 17.20 & \textbf{733.90} & \textbf{750.59} & 6.90 \\
   E9 & 446.49 & 18.95 & 1131.70 & -- & -- & -- & -- & 3575.06 & 926.38 & 1011.13 & 49.52 & 16.38 & \textbf{893.35} & \textbf{916.56} & 14.53 \\
   E10 & 455.07 & 11.58 & 1053.60 & -- & -- & -- & -- & 4197.26 & 874.51 & 931.44 & 35.06 & 15.99 & \textbf{822.85} & \textbf{841.57} & 9.28 \\
   \midrule
   Avg & 429.98 & 17.48 & 1057.62 & -- & -- & -- & -- & 3910.57 & 863.74 & 937.04 & 48.80 & 16.87 & \textbf{838.67} & \textbf{857.59} & 15.62 \\
   \midrule
   F1 & 548.88 & 936.12 & 1721.40 & -- & -- & -- & -- & 6765.23 & 2557.06 & 2681.11 & 63.85 & 124.35 & \textbf{1311.10} & \textbf{1351.20} & 22.09 \\
   F2 & 543.32 & 957.18 & 1763.80 & -- & -- & -- & -- & 4389.44 & 2544.32 & 2628.75 & 30.43 & 121.74 & \textbf{1298.31} & \textbf{1337.41} & 20.01 \\
   F3 & 547.64 & 914.00 & 1549.60 & -- & -- & -- & -- & 5325.04 & 2433.87 & 2473.77 & 16.10 & 116.49 & \textbf{1215.96} & \textbf{1272.23} & 56.87 \\
   F4 & 531.84 & 1391.01 & 1420.40 & -- & -- & -- & -- & 2650.15 & 2116.80 & 2125.39 & 3.77 & 135.95 & \textbf{1100.66} & \textbf{1134.66} & 35.94 \\
   F5 & 538.14 & 1205.69 & 1701.90 & -- & -- & -- & -- & 2502.22 & 2669.68 & 2697.29 & 6.50 & 119.79 & \textbf{1298.55} & \textbf{1331.09} & 21.10 \\
   F6 & 518.47 & 880.31 & 1639.70 & -- & -- & -- & -- & 9663.74 & 2455.95 & 2504.51 & 30.34 & 109.64 & \textbf{1292.52} & \textbf{1368.41} & 83.02 \\
   F7 & 512.98 & 1362.58 & 1384.30 & -- & -- & -- & -- & 2337.04 & 2253.09 & 2253.09 & 0.00 & 120.48 & \textbf{1084.57} & \textbf{1125.37} & 17.55 \\
   F8 & 536.15 & 972.97 & 1544.60 & -- & -- & -- & -- & 5802.21 & 2321.96 & 2374.64 & 26.50 & 107.70 & \textbf{1123.22} & \textbf{1140.42} & 9.52 \\
   F9 & 543.16 & 1729.03 & 1572.90 & -- & -- & -- & -- & 4297.17 & 2510.01 & 2529.09 & 8.81 & 125.42 & \textbf{1263.19} & \textbf{1344.62} & 84.05 \\
   F10 & 546.84 & 1396.28 & 1581.00 & -- & -- & -- & -- & 2593.33 & 2686.98 & 2709.23 & 7.75 & 119.55 & \textbf{1383.08} & \textbf{1419.76} & 15.41 \\
   \midrule
   Avg & 536.74 & 1174.52 & 1587.96 & -- & -- & -- & -- & 4632.56 & 2454.97 & 2497.69 & 19.40 & 120.11 & \textbf{1237.12} & \textbf{1282.52} & 36.56 \\
   \midrule
   G1 & 612.37 & 7581.60 & 2248.00 & -- & -- & -- & -- & 3569.71 & 5089.92 & 5089.92 & 0.00 & 439.40 & \textbf{1744.14} & \textbf{1824.34} & 96.54 \\
   G2 & 605.84 & 7581.60 & 2316.10 & -- & -- & -- & -- & 3926.59 & 10299.50 & 10299.50 & 0.00 & 519.49 & \textbf{1709.70} & \textbf{1799.78} & 78.71 \\
   G3 & 614.20 & 7525.79 & 1885.30 & -- & -- & -- & -- & 3477.90 & 3152.55 & 3152.55 & 0.00 & 461.61 & \textbf{1464.69} & \textbf{1511.86} & 50.36 \\
   G4 & 604.30 & 7581.60 & 2023.20 & -- & -- & -- & -- & 3858.37 & 3277.34 & 3277.34 & 0.00 & 529.01 & \textbf{1508.94} & \textbf{1569.01} & 76.22 \\
   G5 & 633.66 & 7581.60 & 2247.60 & -- & -- & -- & -- & 3502.11 & 3584.11 & 3584.11 & 0.00 & 466.62 & \textbf{1652.88} & \textbf{1681.01} & 17.35 \\
   G6 & 621.46 & 7581.60 & 2144.40 & -- & -- & -- & -- & 3939.74 & 3498.75 & 3498.75 & 0.00 & 570.54 & \textbf{1681.64} & \textbf{1719.18} & 52.60 \\
   G7 & 602.42 & 7301.50 & 1971.50 & -- & -- & -- & -- & 3573.50 & 3214.17 & 3214.17 & 0.00 & 522.37 & \textbf{1536.00} & \textbf{1604.96} & 35.02 \\
   G8 & 618.74 & 7581.60 & 1987.40 & -- & -- & -- & -- & 3459.40 & 3241.68 & 3241.68 & 0.00 & 531.73 & \textbf{1498.38} & \textbf{1535.90} & 53.54 \\
   G9 & 662.70 & 7395.22 & 2415.50 & -- & -- & -- & -- & 3482.24 & 3673.47 & 3673.47 & 0.00 & 446.57 & \textbf{1850.07} & \textbf{1976.27} & 111.56 \\
   G10 & 633.76 & 7374.16 & 2373.40 & -- & -- & -- & -- & 3436.26 & 3751.71 & 3751.71 & 0.00 & 482.81 & \textbf{1785.37} & \textbf{1868.56} & 102.21 \\
   Avg & 620.94 & 7508.63 & 2161.24 & -- & -- & -- & -- & 3622.58 & 4278.32 & 4278.32 & 0.00 & 497.01 & \textbf{1643.18} & \textbf{1709.09} & 67.41 \\
   \midrule
   \multicolumn{2}{c}{Avg time (sec.)} & \multicolumn{2}{c}{4027.59} & \multicolumn{4}{c}{147.27} & \multicolumn{4}{c}{1993.37} & \multicolumn{4}{c}{92.29} \\
   \multicolumn{2}{c}{\# Best} & \multicolumn{2}{c}{11} & \multicolumn{4}{c}{19} & \multicolumn{4}{c}{33} & \multicolumn{4}{c}{62} \\
   \multicolumn{2}{c}{\# Best avg} & \multicolumn{2}{c}{--} & \multicolumn{4}{c}{2} & \multicolumn{4}{c}{8} & \multicolumn{4}{c}{70} \\
\end{longtable}
\egroup

Observe that not all authors provide solutions to all the instances for the \citet{mankowska2014} benchmark dataset. The missing results are marked with a dash. \citet{lasfargeas2019} argued that their solution method could not produce results for instances larger than 75 patients unless some of the problem's constraints are relaxed. We conjecture that this issue is related to the failure of their initial solution heuristic to produce feasible solutions. \citet{neto2019} justified that instances with 200 and 300 patients cannot be solved with the fix-and-optimize matheuristic due to the memory required to keep the MIP model. For instances larger than 200 patients, we can expect only the AVNS of \citet{mankowska2014} and our BRKGA-MP-IPR to produce feasible solutions, with an advantage for our metaheuristic algorithm both terms of solution value and processing time.

\color{black}
\citet{mankowska2014}  provided optimal solutions for subset A, and all four approaches achieved these solutions. In subset B, \citet{lasfargeas2019} individually produced 4 out of 10 best solutions, and \citet{neto2019} produced the best values for the entire instance subset. Our metaheuristic algorithm also provides the four best results for this subset. For the instances with 25 patients, our BRKGA is, on average, the fastest method. It is followed by the VNS from \citet{lasfargeas2019} and by the \textit{matheuristic} from \citet{neto2019}. Our metaheuristic algorithm produces almost all the best results for subset D and greater. Additionally, the metaheuristic algorithm of \citet{mankowska2014} was the fastest method for solving up to 75 patients, but the solutions produced are the worst among all the compared methods. For subsets F and G, the largest test cases, the BRKGA outperformed all the others in both solution quality and runtime.
\normalcolor

\subsection{Comparison with the standard BRKGA}
\label{subsection:comparison-lit-gecco}

\textcolor{black}{As we mentioned in Section~\ref{section:proposed-method}, \citet{kummer2020} tested a very comparable but more limited version of our heuristic consisting of a single population variant with the standard two-parent mating with the explicit setting of bias weight towards the elite parent. Furthermore, their proposed decoder has no extra components, namely the \textit{convex hull} strategy and the \textit{workload balancing heuristic} we proposed in this paper. Their algorithm also only selects candidate vehicles that strictly improve the cost during the greedy construction phase of the decoder.} Despite these differences, both have so many similarities that it makes sense to run a dedicated comparison between the two methods.


\color{black}
Table \ref{table:comparison-literature-gecco2} presents results for each instance from \citet{mankowska2014} dataset (column ``Instance'').
Under ``KBA2020'', we present the average solve time in seconds, the value of the best solution, and the average and standard deviation of solution values produced by the algorithm of \citet{kummer2020}. We run their genetic algorithm 20 times, using the numbers between 1 and 20 as seeds. We present results for two variations of our BRKGA.
The variation ``BRKGA (SD)'' uses the intensification components proposed by \citet{andrade2021} but uses the simpler decoding (SD) algorithm proposed by \citet{kummer2020}. The variation ``BRKGA (FD)'' uses the full decoding (FD) scheme described in Subsection~\ref{subsection:decoding}. For both variations, we report the average number of generations evolved by the algorithms (Gens (avg)), the average solve times (Time (sec)), the best solution value among the 20 runs of each instance (Best), and the average (Avg) and standard deviation (SD) of solution values per instance, respectively. At the bottom, we present the average solve time of each method (Avg time (sec)), the number of instances in which each method produced the best solution value (\#~Best), and the number of instances in which each method produced the best average solution value (\#~Best avg).
\normalcolor

\newpage
\bgroup
\scriptsize
\renewcommand{\tabcolsep}{4pt}
\renewcommand{\arraystretch}{0.94}
\color{black}
\begin{longtable}{lrrrrrrrrrrrrrr}
   \caption{Comparison of \citet{kummer2020} genetic algorithm and our proposed BRKGA.}
   \label{table:comparison-literature-gecco2}\\
   \toprule
   \multirow{2}[5]{*}{Instance} &
   \multicolumn{4}{c}{KBA2020} &
   \multicolumn{5}{c}{BRKGA (SD)} &
   \multicolumn{5}{c}{BRKGA (FD)}\\
   \cmidrule(r){2-5}
   \cmidrule(r){6-10}
   \cmidrule(){11-15}
   &
   \makecell{Time\\(sec)} & Best & Avg & SD &
   \makecell{Gens\\(avg)} & \makecell{Time\\(sec)} & Best & Avg & SD &
   \makecell{Gens\\(avg)} & \makecell{Time\\(sec)} & Best & Avg & SD\\
   \midrule
   \endfirsthead

   \caption*{Table \ref{table:comparison-literature-gecco2}: (Continued) Comparison of \citet{kummer2020} genetic algorithm and our proposed BRKGA.}\\
   \toprule
   \multirow{2}[5]{*}{Instance} &
   \multicolumn{4}{c}{KBA2020} &
   \multicolumn{5}{c}{BRKGA (SD)} &
   \multicolumn{5}{c}{BRKGA (FD)}\\
   \cmidrule(r){2-5}
   \cmidrule(r){6-10}
   \cmidrule(){11-15}
   &
   \makecell{Time\\(sec)} & Best & Avg & SD &
   \makecell{Gens\\(avg)} & \makecell{Time\\(sec)} & Best & Avg & SD &
   \makecell{Gens\\(avg)} & \makecell{Time\\(sec)} & Best & Avg & SD\\
   \midrule
   \endhead

   \midrule
   \endfoot

   \bottomrule
   \endlastfoot

   A1 & 0.77 & 226.98 & 226.98 & 0.00 & 11.05 & 0.20 & 226.98 & 226.98 & 0.00 & 11.25 & 0.20 & \textbf{218.20} & \textbf{218.20} & 0.00 \\
   A2 & 0.77 & \textbf{246.63} & \textbf{246.63} & 0.00 & 9.00 & 0.15 & \textbf{246.63} & \textbf{246.63} & 0.00 & 10.40 & 0.17 & \textbf{246.63} & \textbf{246.63} & 0.00 \\
   A3 & 0.77 & \textbf{305.86} & \textbf{305.86} & 0.00 & 9.05 & 0.18 & \textbf{305.86} & \textbf{305.86} & 0.00 & 9.40 & 0.16 & \textbf{305.86} & \textbf{305.86} & 0.00 \\
   A4 & 0.75 & \textbf{186.90} & \textbf{186.90} & 0.00 & 9.75 & 0.17 & \textbf{186.90} & \textbf{186.90} & 0.00 & 9.75 & 0.16 & \textbf{186.90} & \textbf{186.90} & 0.00 \\
   A5 & 0.82 & 191.97 & 191.97 & 0.00 & 8.45 & 0.14 & 191.97 & 191.98 & 0.01 & 10.10 & 0.18 & \textbf{189.54} & \textbf{189.55} & 0.01 \\
   A6 & 0.79 & \textbf{200.13} & \textbf{200.13} & 0.00 & 8.30 & 0.12 & \textbf{200.13} & \textbf{200.15} & 0.01 & 8.70 & 0.15 & \textbf{200.13} & \textbf{200.13} & 0.01 \\
   A7 & 0.73 & \textbf{225.37} & \textbf{225.37} & 0.00 & 10.90 & 0.19 & \textbf{225.37} & \textbf{225.37} & 0.00 & 10.65 & 0.18 & \textbf{225.37} & \textbf{225.37} & 0.00 \\
   A8 & 0.73 & \textbf{232.05} & \textbf{232.05} & 0.00 & 10.05 & 0.17 & \textbf{232.05} & \textbf{232.05} & 0.00 & 10.00 & 0.16 & \textbf{232.05} & \textbf{232.05} & 0.00 \\
   A9 & 0.83 & 234.21 & 234.21 & 0.00 & 10.60 & 0.17 & 234.21 & 234.37 & 0.70 & 12.20 & 0.20 & \textbf{222.30} & \textbf{223.75} & 3.57 \\
   A10 & 0.73 & \textbf{225.01} & \textbf{225.01} & 0.00 & 7.35 & 0.12 & \textbf{225.01} & \textbf{225.01} & 0.00 & 7.55 & 0.13 & \textbf{225.01} & \textbf{225.01} & 0.00 \\
   \midrule
   Avg & 0.77 & 227.51 & 227.51 & 0.00 & 9.45 & 0.16 & 227.51 & 227.53 & 0.07 & 10.00 & 0.17 & \textbf{225.20} & \textbf{225.35} & 0.36 \\
   \midrule
   B1 & 2.18 & \textbf{428.10} & \textbf{428.32} & 0.25 & 42.65 & 0.74 & \textbf{428.10} & 428.97 & 1.30 & 42.80 & 0.78 & \textbf{428.10} & 428.53 & 0.15 \\
   B2 & 1.99 & 483.63 & 485.31 & 1.54 & 47.30 & 0.82 & 483.63 & 484.26 & 1.04 & 48.35 & 0.87 & \textbf{476.05} & \textbf{476.92} & 2.34 \\
   B3 & 2.23 & \textbf{402.80} & \textbf{402.80} & 0.00 & 51.55 & 0.87 & \textbf{402.80} & 408.21 & 4.41 & 55.05 & 0.99 & \textbf{402.80} & 409.29 & 4.78 \\
   B4 & 2.09 & \textbf{420.29} & 432.55 & 3.37 & 60.00 & 1.02 & 432.16 & 439.93 & 5.85 & 62.35 & 1.13 & 422.06 & \textbf{430.46} & 8.81 \\
   B5 & 1.94 & 372.16 & \textbf{374.65} & 3.09 & 51.30 & 0.89 & 372.16 & 378.57 & 2.45 & 54.95 & 0.98 & \textbf{369.44} & 375.15 & 3.88 \\
   B6 & 2.29 & 471.00 & 471.95 & 1.43 & 55.45 & 0.98 & 471.17 & 474.22 & 1.99 & 66.40 & 1.22 & \textbf{470.59} & \textbf{470.70} & 0.16 \\
   B7 & 2.36 & \textbf{328.67} & \textbf{328.67} & 0.00 & 49.00 & 0.86 & \textbf{328.67} & \textbf{328.71} & 0.19 & 51.85 & 0.93 & \textbf{328.67} & \textbf{328.67} & 0.00 \\
   B8 & 2.35 & 359.70 & 359.70 & 0.00 & 34.15 & 0.59 & 359.70 & 359.70 & 0.00 & 38.00 & 0.70 & \textbf{357.68} & \textbf{359.40} & 0.66 \\
   B9 & 2.49 & \textbf{402.67} & 404.25 & 1.06 & 50.15 & 0.88 & \textbf{402.67} & \textbf{404.10} & 0.54 & 48.65 & 0.91 & 404.11 & 404.29 & 0.23 \\
   B10 & 2.21 & \textbf{469.58} & \textbf{469.58} & 0.00 & 45.15 & 0.79 & \textbf{469.58} & \textbf{469.58} & 0.00 & 46.40 & 0.85 & \textbf{469.58} & \textbf{469.58} & 0.00 \\
   \midrule
   Avg & 2.21 & 413.86 & 415.78 & 1.07 & 48.67 & 0.84 & 415.06 & 417.63 & 1.78 & 51.48 & 0.93 & \textbf{412.91} & \textbf{415.30} & 2.10 \\
   \midrule
   C1 & 7.08 & \textbf{965.15} & 977.56 & 14.39 & 129.80 & 2.96 & 966.96 & \textbf{972.56} & 6.03 & 131.00 & 3.09 & 969.11 & 973.87 & 3.52 \\
   C2 & 7.39 & 583.39 & 590.45 & 8.53 & 121.20 & 2.84 & \textbf{582.22} & 588.15 & 3.04 & 120.20 & 2.88 & 584.18 & \textbf{587.00} & 3.26 \\
   C3 & 7.14 & \textbf{548.79} & 559.53 & 6.44 & 117.30 & 2.72 & 553.05 & 559.33 & 5.60 & 121.15 & 2.86 & 549.63 & \textbf{552.52} & 3.96 \\
   C4 & 6.74 & \textbf{519.91} & 531.91 & 7.38 & 112.45 & 2.57 & 520.97 & 528.59 & 5.25 & 126.55 & 2.97 & 520.13 & \textbf{524.15} & 2.88 \\
   C5 & 6.01 & 678.49 & 698.14 & 19.54 & 137.80 & 3.02 & 678.49 & 698.34 & 14.10 & 148.90 & 3.35 & \textbf{668.65} & \textbf{685.92} & 13.92 \\
   C6 & 6.99 & \textbf{840.69} & \textbf{845.30} & 4.30 & 121.90 & 2.81 & 842.99 & 846.66 & 1.97 & 123.30 & 2.92 & 841.48 & 846.83 & 2.40 \\
   C7 & 8.28 & 534.01 & 540.42 & 5.33 & 133.30 & 3.17 & \textbf{528.56} & \textbf{537.72} & 3.14 & 136.55 & 3.37 & 533.92 & 541.88 & 5.94 \\
   C8 & 6.82 & 474.55 & 479.75 & 3.60 & 127.75 & 2.87 & \textbf{473.82} & 480.68 & 2.66 & 148.60 & 3.46 & 475.96 & \textbf{478.39} & 2.84 \\
   C9 & 8.40 & \textbf{534.30} & 551.72 & 9.25 & 143.60 & 3.44 & 536.99 & \textbf{548.04} & 6.42 & 139.25 & 3.44 & 545.18 & 558.54 & 15.13 \\
   C10 & 6.61 & 611.25 & 618.83 & 4.36 & 115.20 & 2.63 & 611.67 & 617.06 & 3.36 & 116.85 & 2.73 & \textbf{611.03} & \textbf{614.59} & 3.74 \\
   \midrule
   Avg & 7.14 & \textbf{629.05} & 639.36 & 8.31 & 126.03 & 2.90 & 629.57 & 637.71 & 5.16 & 131.24 & 3.10 & 629.93 & \textbf{636.37} & 5.76 \\
   \midrule
   D1 & 17.67 & \textbf{1186.20} & 1208.19 & 13.98 & 187.25 & 6.03 & 1199.65 & \textbf{1206.88} & 4.90 & 235.60 & 7.99 & 1193.21 & 1215.79 & 11.11 \\
   D2 & 14.38 & 693.28 & 719.52 & 15.00 & 222.25 & 6.74 & 695.81 & 713.60 & 12.03 & 231.85 & 7.24 & \textbf{679.58} & \textbf{695.99} & 11.49 \\
   D3 & 18.48 & \textbf{635.67} & 650.29 & 7.63 & 211.60 & 7.14 & 638.82 & \textbf{644.33} & 3.07 & 238.20 & 8.50 & 644.16 & 650.22 & 3.65 \\
   D4 & 14.46 & 809.45 & 840.61 & 13.72 & 187.95 & 5.74 & 811.33 & 841.19 & 10.31 & 227.25 & 7.22 & \textbf{795.15} & \textbf{827.28} & 12.32 \\
   D5 & 17.81 & \textbf{691.50} & 703.23 & 12.10 & 211.90 & 6.90 & 692.54 & \textbf{699.64} & 5.36 & 228.20 & 7.72 & 693.83 & 702.68 & 3.91 \\
   D6 & 19.30 & 733.67 & 744.99 & 7.32 & 195.75 & 6.75 & \textbf{728.94} & 743.78 & 6.16 & 221.30 & 7.91 & 731.71 & \textbf{743.64} & 5.97 \\
   D7 & 20.51 & 590.64 & 603.65 & 6.42 & 209.70 & 7.44 & 588.73 & 606.52 & 14.61 & 211.50 & 7.83 & \textbf{586.10} & \textbf{597.25} & 5.68 \\
   D8 & 18.66 & 661.78 & 681.10 & 12.47 & 196.65 & 6.61 & 660.33 & 670.59 & 6.85 & 230.85 & 8.10 & \textbf{658.49} & \textbf{669.83} & 7.02 \\
   D9 & 17.71 & 704.63 & 722.09 & 10.35 & 238.30 & 7.78 & 706.44 & 719.93 & 7.67 & 271.60 & 9.15 & \textbf{689.83} & \textbf{710.32} & 10.95 \\
   D10 & 13.37 & 1208.71 & 1294.66 & 66.45 & 200.80 & 5.97 & 1236.76 & 1289.61 & 54.11 & 222.05 & 6.87 & \textbf{1189.32} & \textbf{1280.92} & 62.72 \\
   \midrule
   Avg & 17.23 & 791.55 & 816.83 & 16.54 & 206.22 & 6.71 & 795.94 & 813.61 & 12.51 & 231.84 & 7.85 & \textbf{786.14} & \textbf{809.39} & 13.48 \\
   \midrule
   E1 & 33.89 & 1331.49 & 1352.33 & 10.75 & 277.00 & 12.91 & \textbf{1317.90} & \textbf{1337.99} & 13.61 & 347.55 & 17.19 & 1327.72 & 1340.36 & 8.44 \\
   E2 & 33.01 & 848.08 & 869.45 & 15.00 & 327.80 & 15.08 & 837.54 & \textbf{857.31} & 11.91 & 357.25 & 17.14 & \textbf{829.79} & 865.05 & 45.91 \\
   E3 & 32.17 & 788.03 & 815.31 & 15.16 & 307.55 & 13.85 & \textbf{787.94} & \textbf{805.38} & 10.37 & 351.10 & 16.70 & 789.56 & 806.53 & 17.56 \\
   E4 & 34.99 & \textbf{711.19} & 731.24 & 11.75 & 276.55 & 13.03 & 713.78 & \textbf{725.05} & 5.99 & 292.25 & 14.77 & 723.87 & 728.96 & 3.45 \\
   E5 & 32.18 & 781.50 & 806.81 & 13.73 & 319.35 & 14.51 & 782.54 & \textbf{799.66} & 11.26 & 354.15 & 17.03 & \textbf{780.04} & 817.13 & 34.10 \\
   E6 & 34.38 & 788.08 & 803.28 & 7.86 & 328.20 & 15.40 & 790.51 & 799.31 & 6.18 & 371.35 & 18.28 & \textbf{779.82} & \textbf{793.68} & 8.15 \\
   E7 & 37.90 & 711.11 & 731.83 & 9.38 & 285.95 & 14.40 & 712.56 & 728.04 & 7.11 & 343.10 & 18.02 & \textbf{705.79} & \textbf{715.46} & 7.89 \\
   E8 & 31.83 & 748.48 & 761.06 & 7.40 & 284.55 & 12.80 & 737.01 & 762.65 & 16.46 & 366.10 & 17.20 & \textbf{733.90} & \textbf{750.59} & 6.90 \\
   E9 & 30.96 & 921.78 & 950.89 & 13.90 & 324.55 & 14.47 & 901.83 & 939.67 & 17.90 & 352.20 & 16.38 & \textbf{893.35} & \textbf{916.56} & 14.53 \\
   E10 & 33.61 & 825.24 & 847.64 & 14.48 & 294.65 & 13.65 & \textbf{822.24} & \textbf{840.13} & 8.25 & 327.45 & 15.99 & 822.85 & 841.57 & 9.28 \\
   \midrule
   Avg & 33.49 & 845.50 & 866.98 & 11.94 & 302.62 & 14.01 & 840.39 & 859.52 & 10.90 & 346.25 & 16.87 & \textbf{838.67} & \textbf{857.59} & 15.62 \\
   \newpage
   F1 & 133.86 & 1372.69 & 1421.76 & 17.34 & 686.70 & 91.38 & 1369.44 & 1395.91 & 16.45 & 886.80 & 124.35 & \textbf{1311.10} & \textbf{1351.20} & 22.09 \\
   F2 & 143.70 & 1336.33 & 1383.55 & 23.01 & 855.10 & 117.93 & 1301.28 & 1347.48 & 22.79 & 833.60 & 121.74 & \textbf{1298.31} & \textbf{1337.41} & 20.01 \\
   F3 & 137.42 & 1263.39 & 1288.66 & 14.08 & 668.75 & 90.42 & 1230.40 & \textbf{1248.36} & 10.22 & 814.70 & 116.49 & \textbf{1215.96} & 1272.23 & 56.87 \\
   F4 & 169.13 & 1124.24 & 1145.53 & 10.23 & 653.15 & 103.74 & 1111.23 & \textbf{1127.86} & 10.44 & 811.45 & 135.95 & \textbf{1100.66} & 1134.66 & 35.94 \\
   F5 & 153.59 & 1316.54 & 1361.64 & 20.75 & 720.25 & 105.20 & \textbf{1298.00} & \textbf{1323.14} & 16.49 & 770.95 & 119.79 & 1298.55 & 1331.09 & 21.10 \\
   F6 & 128.41 & 1322.89 & 1369.38 & 25.66 & 755.45 & 96.41 & 1307.26 & \textbf{1346.67} & 38.96 & 807.15 & 109.64 & \textbf{1292.52} & 1368.41 & 83.02 \\
   F7 & 164.91 & 1131.27 & 1163.97 & 17.53 & 645.15 & 99.49 & 1110.86 & 1133.37 & 14.68 & 736.90 & 120.48 & \textbf{1084.57} & \textbf{1125.37} & 17.55 \\
   F8 & 145.91 & 1132.77 & 1165.67 & 16.72 & 695.65 & 96.86 & \textbf{1118.15} & \textbf{1139.21} & 10.36 & 726.10 & 107.70 & 1123.22 & 1140.42 & 9.52 \\
   F9 & 155.50 & 1293.78 & 1347.78 & 22.91 & 657.55 & 97.43 & 1264.42 & \textbf{1305.77} & 21.42 & 802.15 & 125.42 & \textbf{1263.19} & 1344.62 & 84.05 \\
   F10 & 149.67 & 1418.53 & 1446.56 & 15.68 & 737.80 & 105.59 & 1392.71 & 1421.59 & 14.29 & 787.05 & 119.55 & \textbf{1383.08} & \textbf{1419.76} & 15.41 \\
   \midrule
   Avg & 148.21 & 1271.24 & 1309.45 & 18.39 & 707.56 & 100.44 & 1250.38 & \textbf{1278.94} & 17.61 & 797.69 & 120.11 & \textbf{1237.12} & 1282.52 & 36.56 \\
   \midrule
   G1 & 348.03 & 1778.54 & 1851.25 & 32.13 & 1224.20 & 413.30 & 1758.18 & \textbf{1787.46} & 16.29 & 1211.20 & 439.40 & \textbf{1744.14} & 1824.34 & 96.54 \\
   G2 & 388.24 & 1824.74 & 1899.33 & 34.71 & 1404.65 & 508.39 & \textbf{1708.75} & \textbf{1780.34} & 28.55 & 1316.45 & 519.49 & 1709.70 & 1799.78 & 78.71 \\
   G3 & 340.51 & 1514.23 & 1546.44 & 13.92 & 1131.70 & 381.37 & \textbf{1453.53} & \textbf{1485.42} & 19.77 & 1294.70 & 461.61 & 1464.69 & 1511.86 & 50.36 \\
   G4 & 397.61 & 1564.42 & 1599.35 & 23.92 & 1113.75 & 417.95 & \textbf{1491.44} & \textbf{1541.94} & 24.76 & 1313.95 & 529.01 & 1508.94 & 1569.01 & 76.22 \\
   G5 & 328.49 & 1694.50 & 1750.03 & 28.07 & 1092.55 & 354.84 & \textbf{1630.25} & \textbf{1671.12} & 23.07 & 1358.10 & 466.62 & 1652.88 & 1681.01 & 17.35 \\
   G6 & 420.99 & 1714.38 & 1779.29 & 21.68 & 1161.00 & 454.69 & \textbf{1671.76} & \textbf{1704.90} & 15.71 & 1353.85 & 570.54 & 1681.64 & 1719.18 & 52.60 \\
   G7 & 354.21 & 1640.07 & 1677.66 & 22.99 & 1203.90 & 411.84 & 1568.52 & 1614.15 & 19.17 & 1430.45 & 522.37 & \textbf{1536.00} & \textbf{1604.96} & 35.02 \\
   G8 & 348.98 & 1547.63 & 1582.71 & 20.05 & 1244.15 & 418.63 & 1499.45 & \textbf{1522.14} & 15.26 & 1486.25 & 531.73 & \textbf{1498.38} & 1535.90 & 53.54 \\
   G9 & 357.05 & 1942.21 & 1974.16 & 19.69 & 1366.90 & 472.92 & \textbf{1836.61} & \textbf{1895.25} & 62.12 & 1190.35 & 446.57 & 1850.07 & 1976.27 & 111.56 \\
   G10 & 317.27 & 1872.08 & 1931.99 & 25.31 & 1283.50 & 403.33 & 1821.23 & \textbf{1863.12} & 55.45 & 1437.50 & 482.81 & \textbf{1785.37} & 1868.56 & 102.21 \\
   \midrule
   Avg & 360.14 & 1709.28 & 1759.22 & 24.25 & 1222.63 & 423.72 & 1643.97 & \textbf{1686.59} & 28.02 & 1339.28 & 497.01 & \textbf{1643.18} & 1709.09 & 67.41 \\
   \midrule
   \makecell[l]{Avg\\time (sec)} & \multicolumn{4}{c}{81.31} & \multicolumn{5}{c}{78.40} & \multicolumn{5}{c}{92.29} \\
   \# Best & \multicolumn{4}{c}{22} & \multicolumn{5}{c}{27} & \multicolumn{5}{c}{44} \\
   \# Best avg & \multicolumn{4}{c}{13} & \multicolumn{5}{c}{37} & \multicolumn{5}{c}{38} \\
\end{longtable}
\egroup

\color{black}
The results indicate a mixed performance among BRKGA (SD) and BRKGA (FD). Very few instances presented better solution via the genetic algorithm of \citet{kummer2020} (B4, C1, C6, D1, D3, D5, and E4), although the largest difference was 1.36\%  between the best solution our algorithm produced to the test case D3 and KBA2020. Furthermore, we credit this difference to the stopping criteria of the solution methods. While KBA2020 used a fixed number of generations on their tests, our algorithm calculates its stopping criterion according to the instance sizes. For instance D1, KBA2020 evolved a solution by 1823 generations, while our BRKGA (FD) achieved, on average, 235.60 generations.
\normalcolor


\color{black}
In general, our BRKGA produced the best solution up to 44 out of 70 instances from the \citet{mankowska2014} dataset, compared to the best 22 results from \citet{kummer2020}. (\#~Best at the bottom of the table). Furthermore, the BRKGA (SD) was the fastest among the three approaches, which indicates that a clever stopping criterion can leverage short computational times, even in the presence of potentially expensive components such as the multi-population and the IPR heuristic. Note also that BRKGA (FD) is, on average, slightly slower than the other two algorithms. This result comes from the larger effort by BRKGA (FD) to converge the solution, mainly due to the increased number of possible solutions that it can achieve. This increased effort  can be further verified in the average number of generations evolved by BRKGA (FD) for subset G (1339.28 versus 1222.63 generations by BRKGA (SD)).
\normalcolor

\color{black}
\subsection{Component analysis for the new genetic algorithm}

Although we already verified the superiority of our new BRKGA in producing better solution values to the HHCRSP, it is still interesting to further analyze by how much each of the additional intensification components proposed by \citet{andrade2021} effectively contributes in producing new solution values. With that in mind, we dissected our algorithm BRKGA (FD) in four variants as follows. Note that we used the best configuration of \textit{irace} from Subsection \ref{subsection:irace} among all these four algorithms. Thus some parameters were just ignored when their relative components were explicit turned off.

\begin{itemize}
   \item BRKGA-MP (FD). Uses the decoding algorithm from Subsection \ref{subsection:decoding}, but only the multi-parent mating component of \citet{andrade2021};
   \item BRKGA-MP-MI (FD). Similar to BRKGA-MP (FD), but also uses the multiple islands to evolve independent populations;
   \item BRKGA-MP-IPR (FD). Similar to BRKGA-MP (FD), but also uses the \textit{implicit path-relinking} of \citet{andrade2021};
   \item BRKGA-MP-MI-IPR (FD). Uses all the intensification components proposed by \citet{andrade2021}, plus the \textit{full decoding} scheme of Subsection \ref{subsection:decoding}. This is the same as ``BRKGA (FD)'' we mention in Subsections \ref{subsection:comparison-lit-ls} and \ref{subsection:comparison-lit-gecco}.
\end{itemize}

Table \ref{table:brkga-component-analysis} lists the extensive computational results for all the four combinations of the intensification components tested. For each approach, the table indicates the average solve time in seconds (Time (sec)), the best solution produced per instance (Best), and the average solution value (Avg). We used the same testing protocol as our other experiments, so we solved each instance 20 times by each approach, using the numbers 1 to 20 as seeds. At the bottom of the table, we summarize some data regarding the whole experiment. We present the average solve times in second (Avg time (sec)), the number of instances in which each approach produced the best solution values (\#~Best), and the number of times which each algorithm produced the best average solution values (\#~Best avg).

\bgroup
\scriptsize
\renewcommand{\tabcolsep}{4pt}
\renewcommand{\arraystretch}{0.95}
\color{black}
\begin{longtable}{lrrrrrrrrrrrr}

   \caption{Results of combined intensification components from the literature.}
   \label{table:brkga-component-analysis}\\
   \toprule
   \multirow{2}[2]{*}{Instance} &
   \multicolumn{3}{c}{BRKGA-MP (FD)} &
   \multicolumn{3}{c}{BRKGA-MP-MI (FD)} &
   \multicolumn{3}{c}{BRKGA-MP-IPR (FD)} &
   \multicolumn{3}{c}{BRKGA-MP-MI-IPR (FD)} \\
   \cmidrule(r){2-4}
   \cmidrule(r){5-7}
   \cmidrule(r){8-10}
   \cmidrule(){11-13}
   & \makecell{Time (sec)} & Best & Avg
   & \makecell{Time (sec)} & Best & Avg
   & \makecell{Time (sec)} & Best & Avg
   & \makecell{Time (sec)} & Best & Avg\\
   \midrule
   \endfirsthead

   \caption*{Table \ref{table:brkga-component-analysis}: (Continued) Results of combined intensification components from the literature.}\\
   \toprule
   \multirow{2}[2]{*}{Instance} &
   \multicolumn{3}{c}{BRKGA-MP (FD)} &
   \multicolumn{3}{c}{BRKGA-MP-MI (FD)} &
   \multicolumn{3}{c}{BRKGA-MP-IPR (FD)} &
   \multicolumn{3}{c}{BRKGA-MP-MI-IPR (FD)} \\
   \cmidrule(r){2-4}
   \cmidrule(r){5-7}
   \cmidrule(r){8-10}
   \cmidrule(){11-13}
   & \makecell{Time (sec)} & Best & Avg
   & \makecell{Time (sec)} & Best & Avg
   & \makecell{Time (sec)} & Best & Avg
   & \makecell{Time (sec)} & Best & Avg\\
   \midrule
   \endhead

   \midrule
   \endfoot

   \bottomrule
   \endlastfoot

   A1 & 0.11 & \textbf{218.20} & 218.33 & 0.20 & \textbf{218.20} & \textbf{218.20} & 0.10 & \textbf{218.20} & 218.33 & 0.20 & \textbf{218.20} & \textbf{218.20} \\
   A2 & 0.10 & \textbf{246.63} & \textbf{246.63} & 0.18 & \textbf{246.63} & \textbf{246.63} & 0.10 & \textbf{246.63} & \textbf{246.63} & 0.17 & \textbf{246.63} & \textbf{246.63} \\
   A3 & 0.10 & \textbf{305.86} & \textbf{305.86} & 0.15 & \textbf{305.86} & \textbf{305.86} & 0.10 & \textbf{305.86} & \textbf{305.86} & 0.16 & \textbf{305.86} & \textbf{305.86} \\
   A4 & 0.10 & \textbf{186.90} & \textbf{186.90} & 0.16 & \textbf{186.90} & \textbf{186.90} & 0.10 & \textbf{186.90} & \textbf{186.90} & 0.16 & \textbf{186.90} & \textbf{186.90} \\
   A5 & 0.10 & \textbf{189.54} & \textbf{189.55} & 0.17 & \textbf{189.54} & \textbf{189.55} & 0.10 & \textbf{189.54} & \textbf{189.55} & 0.18 & \textbf{189.54} & \textbf{189.55} \\
   A6 & 0.10 & \textbf{200.13} & \textbf{200.14} & 0.15 & \textbf{200.13} & \textbf{200.13} & 0.10 & \textbf{200.13} & \textbf{200.14} & 0.15 & \textbf{200.13} & \textbf{200.13} \\
   A7 & 0.10 & \textbf{225.37} & \textbf{225.37} & 0.18 & \textbf{225.37} & \textbf{225.37} & 0.10 & \textbf{225.37} & \textbf{225.37} & 0.18 & \textbf{225.37} & \textbf{225.37} \\
   A8 & 0.10 & \textbf{232.05} & \textbf{232.05} & 0.17 & \textbf{232.05} & \textbf{232.05} & 0.10 & \textbf{232.05} & \textbf{232.05} & 0.16 & \textbf{232.05} & \textbf{232.05} \\
   A9 & 0.14 & \textbf{222.30} & \textbf{223.13} & 0.21 & \textbf{222.30} & 223.75 & 0.13 & \textbf{222.30} & \textbf{223.13} & 0.20 & \textbf{222.30} & 223.75 \\
   A10 & 0.09 & \textbf{225.01} & \textbf{225.01} & 0.12 & \textbf{225.01} & \textbf{225.01} & 0.10 & \textbf{225.01} & \textbf{225.01} & 0.13 & \textbf{225.01} & \textbf{225.01} \\
   \midrule
   Avg & 0.10 & \textbf{225.20} & \textbf{225.30} & 0.17 & \textbf{225.20} & \textbf{225.35} & 0.10 & \textbf{225.20} & \textbf{225.30} & 0.17 & \textbf{225.20} & \textbf{225.35} \\
   \newpage
   B1 & 0.40 & \textbf{428.10} & 429.03 & 0.76 & \textbf{428.10} & \textbf{428.53} & 0.40 & \textbf{428.10} & 429.03 & 0.78 & \textbf{428.10} & \textbf{428.53} \\
   B2 & 0.45 & \textbf{476.05} & 478.53 & 0.86 & \textbf{476.05} & 477.24 & 0.45 & \textbf{476.05} & 478.34 & 0.87 & \textbf{476.05} & \textbf{476.92} \\
   B3 & 0.47 & \textbf{402.80} & 412.30 & 0.95 & \textbf{402.80} & 410.15 & 0.49 & \textbf{402.80} & 412.49 & 0.99 & \textbf{402.80} & \textbf{409.29} \\
   B4 & 0.54 & \textbf{422.06} & 434.86 & 1.12 & \textbf{422.06} & \textbf{429.54} & 0.54 & \textbf{422.06} & 434.06 & 1.13 & \textbf{422.06} & 430.46 \\
   B5 & 0.51 & 370.49 & 376.17 & 0.98 & 370.49 & \textbf{374.63} & 0.50 & \textbf{369.44} & 376.43 & 0.98 & \textbf{369.44} & 375.15 \\
   B6 & 0.59 & \textbf{470.59} & 472.26 & 1.19 & \textbf{470.59} & 470.80 & 0.59 & \textbf{470.59} & 472.25 & 1.22 & \textbf{470.59} & \textbf{470.70} \\
   B7 & 0.48 & \textbf{328.67} & 328.86 & 0.96 & \textbf{328.67} & \textbf{328.67} & 0.48 & \textbf{328.67} & 328.88 & 0.93 & \textbf{328.67} & \textbf{328.67} \\
   B8 & 0.32 & 358.69 & 359.60 & 0.69 & 358.69 & \textbf{359.35} & 0.33 & 358.69 & 359.60 & 0.70 & \textbf{357.68} & 359.40 \\
   B9 & 0.46 & \textbf{404.11} & 404.76 & 0.89 & \textbf{404.11} & \textbf{404.29} & 0.48 & \textbf{404.11} & 404.74 & 0.91 & \textbf{404.11} & \textbf{404.29} \\
   B10 & 0.46 & \textbf{469.58} & \textbf{469.61} & 0.83 & \textbf{469.58} & \textbf{469.58} & 0.47 & \textbf{469.58} & \textbf{469.61} & 0.85 & \textbf{469.58} & \textbf{469.58} \\
   \midrule
   Avg & 0.47 & 413.11 & 416.60 & 0.92 & 413.11 & \textbf{415.28} & 0.47 & 413.01 & 416.54 & 0.93 & \textbf{412.91} & \textbf{415.30} \\
   \midrule
   C1 & 1.45 & 970.36 & 980.74 & 3.24 & 970.36 & \textbf{972.34} & 1.42 & \textbf{969.01} & 980.20 & 3.09 & 969.11 & 973.87 \\
   C2 & 1.48 & 584.25 & 589.52 & 2.83 & 584.25 & 587.42 & 1.49 & 584.25 & 589.73 & 2.88 & \textbf{584.18} & \textbf{587.00} \\
   C3 & 1.48 & \textbf{549.35} & 554.66 & 2.90 & \textbf{549.35} & \textbf{552.26} & 1.46 & 549.63 & 555.08 & 2.86 & 549.63 & 552.52 \\
   C4 & 1.46 & \textbf{519.91} & 527.11 & 2.96 & \textbf{519.91} & 524.83 & 1.54 & \textbf{519.91} & 526.49 & 2.97 & 520.13 & \textbf{524.15} \\
   C5 & 1.60 & 674.69 & 705.11 & 3.09 & 674.69 & 688.88 & 1.61 & 679.88 & 703.80 & 3.35 & \textbf{668.65} & \textbf{685.92} \\
   C6 & 1.44 & 843.73 & 851.38 & 2.72 & 843.73 & 847.77 & 1.53 & 845.82 & 851.77 & 2.92 & \textbf{841.48} & \textbf{846.83} \\
   C7 & 1.66 & \textbf{533.74} & 544.14 & 3.49 & \textbf{533.74} & \textbf{539.72} & 1.69 & 538.03 & 544.46 & 3.37 & 533.92 & 541.88 \\
   C8 & 1.69 & \textbf{475.81} & 478.77 & 3.42 & \textbf{475.81} & 478.44 & 1.63 & 476.20 & 480.19 & 3.46 & 475.96 & \textbf{478.39} \\
   C9 & 1.91 & 538.04 & \textbf{558.05} & 3.33 & 538.04 & 562.49 & 1.75 & \textbf{536.88} & 561.61 & 3.44 & 545.18 & 558.54 \\
   C10 & 1.41 & \textbf{611.03} & 617.76 & 2.77 & \textbf{611.03} & 615.05 & 1.50 & \textbf{611.03} & 617.49 & 2.73 & \textbf{611.03} & \textbf{614.59} \\
   \midrule
   Avg & 1.56 & 630.09 & 640.72 & 3.07 & 630.09 & 636.92 & 1.56 & 631.06 & 641.08 & 3.10 & \textbf{629.93} & \textbf{636.37} \\
   \midrule
   D1 & 3.94 & 1198.16 & 1218.58 & 8.18 & 1198.16 & \textbf{1210.86} & 3.96 & 1199.55 & 1219.44 & 7.99 & \textbf{1193.21} & 1215.79 \\
   D2 & 3.34 & 685.75 & 705.39 & 6.96 & 685.75 & 697.31 & 3.66 & 686.24 & 709.32 & 7.24 & \textbf{679.58} & \textbf{695.99} \\
   D3 & 3.87 & \textbf{638.32} & 652.23 & 8.03 & \textbf{638.32} & \textbf{650.10} & 4.07 & 643.77 & 651.88 & 8.50 & 644.16 & 650.22 \\
   D4 & 3.55 & 802.84 & 834.41 & 6.62 & 802.84 & 829.25 & 3.39 & 805.58 & 836.03 & 7.22 & \textbf{795.15} & \textbf{827.28} \\
   D5 & 3.76 & 693.17 & 708.35 & 7.34 & 693.17 & \textbf{702.23} & 3.70 & \textbf{692.58} & 707.27 & 7.72 & 693.83 & 702.68 \\
   D6 & 4.04 & 735.43 & 748.01 & 7.87 & 735.43 & \textbf{743.51} & 3.85 & 737.07 & 749.53 & 7.91 & \textbf{731.71} & 743.64 \\
   D7 & 3.83 & 590.19 & 602.78 & 8.42 & 590.19 & \textbf{596.59} & 3.76 & 591.90 & 604.78 & 7.83 & \textbf{586.10} & 597.25 \\
   D8 & 3.94 & 663.69 & 677.42 & 7.53 & 663.69 & 670.61 & 3.93 & 666.12 & 681.46 & 8.10 & \textbf{658.49} & \textbf{669.83} \\
   D9 & 4.56 & 700.52 & 720.21 & 8.83 & 700.52 & 715.10 & 4.25 & 693.62 & 722.10 & 9.15 & \textbf{689.83} & \textbf{710.32} \\
   D10 & 3.45 & 1231.59 & 1324.11 & 6.47 & 1231.59 & 1287.24 & 3.10 & 1236.28 & 1330.68 & 6.87 & \textbf{1189.32} & \textbf{1280.92} \\
   \midrule
   Avg & 3.83 & 793.97 & 819.15 & 7.62 & 793.97 & 810.28 & 3.76 & 795.27 & 821.25 & 7.85 & \textbf{786.14} & \textbf{809.39} \\
   \midrule
   E1 & 7.54 & 1336.46 & 1351.90 & 16.90 & 1336.46 & \textbf{1338.61} & 8.58 & 1330.70 & 1348.62 & 17.19 & \textbf{1327.72} & 1340.36 \\
   E2 & 8.92 & \textbf{818.69} & \textbf{853.75} & 18.22 & \textbf{818.69} & 853.93 & 8.66 & 833.48 & 866.48 & 17.14 & 829.79 & 865.05 \\
   E3 & 8.55 & \textbf{784.89} & 808.47 & 18.08 & \textbf{784.89} & \textbf{803.01} & 8.28 & 793.33 & 815.41 & 16.70 & 789.56 & 806.53 \\
   E4 & 7.11 & 719.89 & 729.39 & 13.71 & 719.89 & 729.28 & 7.51 & \textbf{713.59} & \textbf{728.82} & 14.77 & 723.87 & 728.96 \\
   E5 & 8.83 & 787.01 & 810.93 & 17.26 & 787.01 & \textbf{799.44} & 8.95 & 781.20 & 809.80 & 17.03 & \textbf{780.04} & 817.13 \\
   E6 & 8.66 & 784.54 & 800.74 & 17.67 & 784.54 & 800.21 & 9.02 & 786.32 & 802.22 & 18.28 & \textbf{779.82} & \textbf{793.68} \\
   E7 & 8.10 & 706.81 & 725.04 & 16.28 & 706.81 & 716.00 & 8.73 & \textbf{704.42} & 720.59 & 18.02 & 705.79 & \textbf{715.46} \\
   E8 & 7.94 & 737.83 & \textbf{750.62} & 14.25 & 737.83 & 751.94 & 7.73 & 746.26 & 755.86 & 17.20 & \textbf{733.90} & \textbf{750.59} \\
   E9 & 8.27 & 907.58 & 937.97 & 16.42 & 907.58 & 921.63 & 9.53 & 916.17 & 935.03 & 16.38 & \textbf{893.35} & \textbf{916.56} \\
   E10 & 7.77 & 824.25 & 845.67 & 15.28 & 824.25 & 843.52 & 8.12 & 831.13 & 846.80 & 15.99 & \textbf{822.85} & 841.57 \\
   \midrule
   Avg & 8.17 & 840.80 & 861.45 & 16.40 & 840.80 & \textbf{855.76} & 8.51 & 843.66 & 862.97 & 16.87 & \textbf{838.67} & 857.59 \\
   \midrule
   F1 & 56.36 & 1333.99 & 1367.22 & 121.40 & 1333.99 & 1357.88 & 63.80 & 1331.25 & 1372.64 & 124.35 & \textbf{1311.10} & 1351.20 \\
   F2 & 55.02 & 1323.80 & 1345.16 & 117.58 & 1323.80 & \textbf{1337.45} & 67.65 & 1308.09 & 1352.35 & 121.74 & \textbf{1298.31} & 1337.41 \\
   F3 & 54.04 & 1226.14 & 1260.75 & 113.02 & 1226.14 & \textbf{1252.67} & 61.93 & 1227.43 & 1258.71 & 116.49 & \textbf{1215.96} & 1272.23 \\
   F4 & 57.77 & 1108.51 & 1131.85 & 131.34 & 1108.51 & \textbf{1129.72} & 68.72 & \textbf{1094.77} & 1133.73 & 135.95 & 1100.66 & 1134.66 \\
   F5 & 53.86 & 1314.89 & 1343.67 & 121.37 & 1314.89 & 1332.29 & 60.71 & 1326.11 & 1345.59 & 119.79 & \textbf{1298.55} & 1331.09 \\
   F6 & 48.34 & 1311.24 & \textbf{1348.44} & 102.82 & 1311.24 & 1356.85 & 51.12 & 1310.66 & 1366.06 & 109.64 & \textbf{1292.52} & 1368.41 \\
   F7 & 54.66 & 1110.95 & 1137.43 & 113.36 & 1110.95 & 1133.37 & 65.54 & 1104.15 & 1134.54 & 120.48 & \textbf{1084.57} & 1125.37 \\
   F8 & 46.82 & 1127.27 & 1148.04 & 95.26 & 1127.27 & 1140.94 & 55.61 & \textbf{1120.09} & 1150.68 & 107.70 & 1123.22 & 1140.42 \\
   F9 & 55.43 & 1279.43 & 1315.39 & 120.59 & 1279.43 & \textbf{1313.67} & 66.08 & 1293.61 & 1320.42 & 125.42 & \textbf{1263.19} & 1344.62 \\
   F10 & 51.70 & \textbf{1374.92} & 1429.27 & 102.23 & \textbf{1374.92} & 1425.94 & 60.60 & 1386.55 & 1424.99 & 119.55 & 1383.08 & 1419.76 \\
   \midrule
   Avg & 53.40 & 1251.11 & 1282.72 & 113.89 & 1251.11 & \textbf{1278.08} & 62.17 & 1250.27 & 1285.97 & 120.11 & \textbf{1237.12} & 1282.52 \\
   \newpage
   G1 & 215.48 & 1759.49 & 1797.37 & 465.45 & 1759.49 & \textbf{1771.01} & 265.46 & 1760.69 & 1813.14 & 439.40 & \textbf{1744.14} & 1824.34 \\
   G2 & 250.01 & 1736.98 & 1787.91 & 499.88 & 1736.98 & \textbf{1773.96} & 319.97 & 1738.21 & 1799.39 & 519.49 & \textbf{1709.70} & 1799.78 \\
   G3 & 189.77 & 1480.51 & 1510.20 & 457.79 & 1480.51 & \textbf{1487.32} & 253.78 & 1474.21 & 1510.04 & 461.61 & \textbf{1464.69} & 1511.86 \\
   G4 & 221.62 & 1513.00 & 1553.84 & 496.57 & 1513.00 & \textbf{1548.89} & 321.27 & 1511.15 & 1549.92 & 529.01 & \textbf{1508.94} & 1569.01 \\
   G5 & 199.71 & 1671.42 & 1701.86 & 423.41 & 1671.42 & \textbf{1680.65} & 252.90 & 1658.72 & 1710.80 & 466.62 & \textbf{1652.88} & 1681.01 \\
   G6 & 224.13 & 1700.53 & 1728.20 & 501.02 & 1700.53 & \textbf{1716.80} & 308.38 & 1698.21 & 1738.29 & 570.54 & \textbf{1681.64} & 1719.18 \\
   G7 & 207.47 & 1584.82 & 1612.97 & 451.29 & 1584.82 & \textbf{1591.85} & 302.10 & 1591.10 & 1606.23 & 522.37 & \textbf{1536.00} & 1604.96 \\
   G8 & 206.47 & 1515.08 & 1548.83 & 431.53 & 1515.08 & \textbf{1530.88} & 291.82 & 1508.53 & 1538.69 & 531.73 & \textbf{1498.38} & 1535.90 \\
   G9 & 258.00 & \textbf{1844.29} & 1890.54 & 552.71 & \textbf{1844.29} & \textbf{1883.73} & 272.46 & 1862.58 & 1943.51 & 446.57 & 1850.07 & 1976.27 \\
   G10 & 214.86 & 1828.34 & 1865.95 & 469.80 & 1828.34 & \textbf{1850.58} & 263.73 & 1790.54 & 1892.33 & 482.81 & \textbf{1785.37} & 1868.56 \\
   \midrule
   Avg & 218.75 & 1663.45 & 1699.77 & 474.94 & 1663.45 & \textbf{1683.57} & 285.18 & 1659.39 & 1710.23 & 497.01 & \textbf{1643.18} & 1709.09 \\
   \midrule
   \makecell[l]{Avg\\time (sec)} & \multicolumn{3}{c}{40.89} & \multicolumn{3}{c}{88.15} & \multicolumn{3}{c}{51.68} & \multicolumn{3}{c}{92.29} \\
   \# Best & \multicolumn{3}{c}{28} & \multicolumn{3}{c}{28} & \multicolumn{3}{c}{28} & \multicolumn{3}{c}{54} \\
   \# Best avg & \multicolumn{3}{c}{14} & \multicolumn{3}{c}{41} & \multicolumn{3}{c}{11} & \multicolumn{3}{c}{38} \\
\end{longtable}
\egroup

The first thing to notice is that we can sort the algorithms by their average solve times. The BRKGA-MP (FD) was the fastest approach, requiring on average 218.75 seconds to finish on the largest test cases with 300 patients (subset G). This algorithm is 60.7\% faster compared to KBA2020 (c.f. Table \ref{table:comparison-literature-gecco2}). BRKGA-MP-IPR was the next fastest algorithm, which indicates a relatively small overhead associated with the IPR heuristic, even though the configuration used with the algorithms triggered the IPR quite frequently, between a large number of guide and base solutions (c.f. Subsection \ref{subsection:irace}). It is easy to see that multi-population is the most expensive feature within all four intensification components. Compared to BRKGA-MP (FD), the BRKGA-MP-MI (FD) consumed 2.17 more time to finish, mostly because this algorithm had to evolve two isolated populations in parallel; thus, we expect an increase in algorithm runtime by a factor of two, plus some overhead due to the immigration process.

All the three algorithms discussed so far have similar performance, producing the best solution values for 28 out of 70 instances of the \citet{mankowska2014} dataset--although the best solution by each algorithm refers to distinct instances. Compared to the other algorithms, the slowest yet best-performing BRKGA-MP-MI-IPR (FD) produced the best solutions to 54 out of 70 test cases, which is almost twice the number of best solutions provided by any other algorithm. We can credit its best performance to the IPR heuristic, which seems to perform much better in the presence of the multi-population component; thus, it allows the heuristic to trigger the \textit{implicit path-relinking} between elite individuals from distinct islands. Again, it also has similar solve times to BRKGA-MP-MI (FD), mainly due to the additional overhead of evolving two isolated populations. Both BRKGA-MP-MI (FD) and BRKGA-MP-MI-IPR (FD) have a similar performance regarding \#~Best~avg, but the first algorithm showed a better performance in producing the best average solution values for the subset G, which indicates better stability of results for BRKGA-MP-MI (FD). We think this can be related to the IPR because it is an intensification mechanism that could cause an early loss in population diversity, especially in larger test cases. Still, this would require additional testing and probably new AAC experiments to produce a tailed configuration for each of the four combinations of the intensification components.

As we have a substantial number of experiments, we applied a pairwise Wilcoxon signed-rank test to identify which of the proposed variations of our metaheuristic algorithm are statistically different. According to Figure \ref{figure:parwise-wilcox-test}, almost all pairs of algorithm variations behave differently, with two notable exceptions. First, the statistical test identified no difference between BRKGA-MP (FD) and BRKGA-MP-IPR (FD), which indicates that the IPR heuristic has a negligible impact in such a variant without the multiple populations. Secondly, the test indicated no statistical difference between BRKGA (FD) and BRKGA-MP-MI (FD). The statistical test used a confidence level o 5\%, and the Bonferroni correction was applied. Roughly speaking, a Wilcoxon test works by comparing the distribution of the solution values by each algorithm, so counting which method produces most of the best solutions per instance has no importance. Nevertheless, the BRKGA (FD) variant still produced the largest number of "best solution" over all the studied variations, as presented in Table \ref{table:brkga-component-analysis}.

\begin{figure}[!htb]
   \begin{mdframed}
      \footnotesize
      \color{black}
      \begin{verbatim}
> with(AllResults, pairwise.wilcox.test(x=cost, g=algorithm, p.adj="bonf", paired=T))

Pairwise comparisons using Wilcoxon signed rank test

data:  cost and algorithm

                  BRKGA (FD) BRKGA (SD) BRKGA-MP (FD) BRKGA-MP-IPR (FD) BRKGA-MP-MI (FD)
BRKGA (SD)        1.8e-06    -          -             -                 -
BRKGA-MP (FD)     < 2e-16    2.9e-08    -             -                 -
BRKGA-MP-IPR (FD) < 2e-16    2.1e-06    1             -                 -
BRKGA-MP-MI (FD)  1          1.4e-07    < 2e-16       < 2e-16           -
KBA2020           < 2e-16    < 2e-16    < 2e-16       < 2e-16           < 2e-16
      \end{verbatim}
   \end{mdframed}
   \caption{Output of pairwise Wilcoxon signed rank test for paired samples.}
   \label{figure:parwise-wilcox-test}
\end{figure}

\normalcolor

\section{Conclusion}
\label{section:conclusion}

In this paper, we propose a new multi-population multi-parent biased random-key genetic algorithm for solving the daily home health care problem. This problem consists of a vehicle
routing problem with time-windows, in which the vehicles represent skilled caregivers, and the nodes represent patients requiring one or more service types. To enable such attendance, a matching of skilled caregivers to patient service type requests is required. Additional route inter-dependency constraints are imposed for patients requiring more than one service type.
Our implementation, and the benchmark dataset are publicly available at \texttt{\url{https://github.com/afkummer/brkga-mp-ipr-hhcrsp-2021}}. This repository also contains additional logfiles regarding our automatic parameter configuration experiment, as well as supplementary material regarding our experiments.

We present a heuristic-powered decoder for the problem, and we use a state-of-art tool to perform the automatic parameter configuration of our proposed metaheuristic algorithm. We use a benchmark dataset from the literature to both configure and benchmark our algorithm. We run an extensive experiment where we compare our algorithm with  previously proposed methods. With respect to  previous \textcolor{black}{local search-based methods, we consistently find improvements up to 26.1\% from previously published results. Compared to the similar work of \citet{kummer2020}, our algorithm finds improvements from around 0.4\% up to 6.36\% in the largest test cases.}

\textcolor{black}{There are a few possible directions for improvement of our algorithms. We would like to improve the heuristic decoder further improve its capability of exploring the problem's solution space.} We also plan to run a more in-depth study of the benchmark dataset characteristics to automatically provide more accurate parameter settings for the metaheuristic through a combined algorithm configuration and selection framework. We are currently in contact with a manager that coordinates a pilot project to implement this type of service in a large Brazilian city. Our goal is to adapt our solution technique to solve a real problem the manager currently deals with every week. As a result of the longer planning period, we also target re-planning the problem due to uncertainties.

\section*{Acknowledgements}

\color{black}
We would like to thank to the reviewers for their valuable considerations. This research has the support of FAPERGS, project PqG 17/2551-0001201-1. The first author would like to thank the Coordination for the Improvement of Higher Education Personnel (CAPES) for his doctoral scholarship. The contributions of Luciana Buriol and Mauricio Resende to this paper are not related to their roles at Amazon.
\normalcolor

\bibliographystyle{itor}
\bibliography{itor-hhcrsp-2021}

\newpage
\appendix

\end{document}